\documentclass[nohyper,letterpaper,11pt,notoc]{JHEP3}
\usepackage{graphicx}
\usepackage{amssymb}
\usepackage{amsmath}
\usepackage{amsfonts}
\usepackage{latexsym}
\usepackage[all]{xy}
\usepackage{slashed}
\preprint{}

\newcommand{\keV}{~\mathrm{keV}}
\newcommand{\MeV}{~\mathrm{MeV}}
\newcommand{\GeV}{~\mathrm{GeV}}
\newcommand{\TeV}{~\mathrm{TeV}}

\newcommand{\gps}{g^{\prime 2}}
\newcommand{\ML}{M_{\scriptscriptstyle \Psi}}
\newcommand{\be}{\begin{eqnarray}}
\newcommand{\ee}{\end{eqnarray}}
\newcommand {\unit} [1] {\; \mathrm {#1}}
\newcommand {\cL} {\mathcal{L}}
\newcommand {\cO} {\mathcal{O}}
\newcommand{\gD}{G_{\rm dark}}

\def\Ptmiss{\not{\hbox{\kern-3pt $E_T$}}}

\title{Non-Abelian Dark Sectors and Their Collider Signatures}
\author{Matthew Baumgart$^{(a,b)}$, Clifford Cheung$^{(a,b)}$, Joshua
  T. Ruderman$^{(c)}$, Lian-Tao Wang$^{(c)}$ and Itay
  Yavin$^{(c)}$\\ \it{(a) School of Natural Sciences, Institute for
    Advanced Study, Princeton, NJ 08540} \\ \it{(b) Department of
    Physics, Harvard University, Cambridge, MA 02138} \\ \it{(c)
    Department of Physics, Princeton University, Princeton, NJ 08544}}

\abstract{ Motivated by the recent proliferation of observed
astrophysical anomalies, Arkani-Hamed $et\ al.$ have proposed a
model in which dark matter is charged under a non-abelian ``dark''
gauge symmetry that is broken at $\sim 1 \unit{GeV}$.
  In this paper, we present a survey of concrete models realizing such a scenario, followed
by a largely model-independent study of collider phenomenology
relevant to the Tevatron and the
  LHC\@.  We address some model building issues that are easily surmounted to accommodate the
astrophysics.  While SUSY is not necessary, we argue that it is
theoretically well-motivated
 because the GeV scale is automatically generated.  Specifically, we propose a
novel mechanism by which mixed D-terms in the dark sector induce
either SUSY breaking or a super-Higgs mechanism precisely at a
GeV\@. Furthermore, we elaborate on the original proposal of
Arkani-Hamed $et\ al.$ in which the dark matter acts as a
messenger of gauge mediation to the dark sector. In our collider
analysis we present cross-sections for dominant production
channels and lifetime estimates for primary decay modes.  We find
that dark gauge bosons can be produced at the Tevatron and the
LHC, either through a process analogous to prompt photon
production or through a rare Z decay channel. Dark gauge bosons
will decay back to the SM via ``lepton jets'' which typically
contain $>2$ and as many as $8$ leptons, significantly improving
their discovery potential. Since SUSY decays from the MSSM will
eventually cascade down to these lepton jets, the discovery
potential for direct electroweak-ino production may also be
improved. Exploiting the unique kinematics, we find that it is
possible to reconstruct the mass of the MSSM LSP\@. We also present
several non-SUSY and SUSY decay channels that have displaced
vertices and lead to multiple leptons with partially correlated
impact parameters.}

\begin{document}
\section{Introduction}

Several intriguing observational results from high-energy
astrophysics have motivated an exciting new proposal
\cite{ArkaniHamed:2008qn} in which a WIMP-like dark matter (DM)
particle at 500-800 GeV annihilates primarily into leptons and is
charged under a new ``dark'' force carrier.  ATIC \cite{Chang:2008zz} detects an abundance of
cosmic ray electrons between $300-800\GeV$, while PAMELA
\cite{Adriani:2008zr} sees an excess of positrons (but not
anti-protons \cite{Adriani:2008zq}) at 10-100 GeV\@. Together with the CMB haze
\cite{Finkbeiner:2003im,Finkbeiner:2004us,Dobler:2007wv}, these observations paint a consistent picture whereby DM annihilates primarily into muons and/or electrons \cite{Cholis:2008wq}.

There are two sources of tension between these results and
more conventional models of WIMP dark matter. First, assuming
thermal freeze-out, the standard relic abundance calculation
implies an annihilation cross-section that is at least a hundred
times too small to explain the lepton excesses observed in
astrophysical experiments. A ``boost factor,'' typically attributed
to local over-densities of dark matter, is often evoked in this
case. A second difficulty is the non-observation of corresponding
excesses in anti-protons \cite{Adriani:2008zq} and gamma rays
\cite{Baltz:2008wd}, which puts strong bounds on hadronic channels
that are present in many dark matter models.

\begin{figure}
\begin{center}
 \includegraphics[scale=0.5]{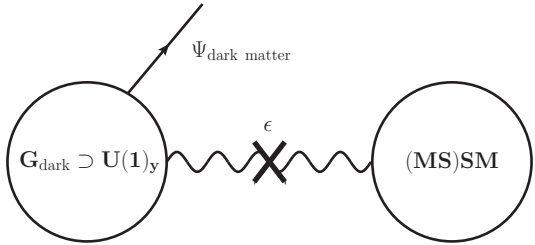}
\end{center}
\caption{A schematic illustration of the minimal setup we consider
in this paper. The dark sector and the SM are connected through
kinetic mixing term suppressed by $\epsilon \lesssim 10^{-3}$. The
dark matter multiplet may or may not couple directly to the SM\@.
Supersymmetric extensions of this scenario are also discussed.}
\label{fig:kineticmixing}
\end{figure}

Motivated by the above considerations, the authors of Ref.~\cite{ArkaniHamed:2008qn} outline a scenario in which these
apparent contradictions are reconciled.  They introduce a 500-800 GeV
WIMP that couples to a GeV scale dark gauge boson that
kinetically mixes with the photon of the Standard Model (SM)
\cite{Holdom:1985ag}\footnote{Ref.~\cite{Pospelov:2008zw} analyzes
particle physics bounds on such a light vector field and its possible
connection to the HyperCP anomaly.} (see Ref. \cite{Pospelov:2007mp} for another recent suggestion with similar ingredients). A schematic illustration of this scenario is presented in Fig.~\ref{fig:kineticmixing}. The ATIC and PAMELA data are
explained by DM annihilation into the dark gauge boson which
subsequently decays into electrons and muons. Elegantly enough,
the $\cO(1)\GeV$ scale plays two independent roles.  First, the
new dark force carrier at $\lesssim\GeV$ introduces a Sommerfeld
enhancement \cite{Hisano:2006nn,Hisano:2004ds,Cirelli:2007xd,MarchRussell:2008yu},
giving a boost factor of the right size to enhance the DM annihilation
cross-section\footnote{See Ref.~\cite{Pospelov:2008jd} for an alternative, but related way for producing a large boost factor.}. Second, the absence of anti-protons in the PAMELA
observations is now simply a result of kinematics
\cite{Cholis:2008vb}.

The gauge group, $G_{\rm dark}$, is a priori unspecified. However,
it was observed in Ref.\ \cite{ArkaniHamed:2008qn} that a non-abelian
$G_{\rm dark}$ nicely accommodates the excited dark matter (XDM)
\cite{Finkbeiner:2007kk} and inelastic dark matter (iDM)
\cite{TuckerSmith:2001hy} mechanisms. XDM was proposed in order to explain the
INTEGRAL~\cite{Weidenspointner:2006nua} measurement of the $511\keV$
gamma-ray line at the center of the galaxy. The iDM scenario can accommodate the DAMA/LIBRA
measurement of WIMP-nuclei scattering with other direct detection experiments
\cite{Akerib:2005kh,Ahmed:2008eu,Angle:2007uj}.  Both XDM and iDM are non-standard
WIMP scenarios in which a DM ground state can transition to
and from new excited states via the emission of some field
that couples back to the SM\@. If the DM lives in a multiplet of a
non-abelian $\gD$, then these ground and excited states can be the
components of this multiplet, and transitions will emit dark gauge
bosons that couple weakly to the SM\@. Independent of the results
from INTEGRAL and DAMA, we find the possibility of a non-abelian
dark sector to be intriguing in its own right, with direct
implications for the collider phenomenology. Thus throughout this
paper we consider a dark sector with a non-abelian gauge symmetry
that is completely broken by some dark
Higgs sector\footnote{There are strong astrophysical
constraints on a long range interaction from unbroken gauge
symmetry with an unsuppressed coupling
\cite{Spergel:1999mh,Dave:2000ar,Ackerman:2008gi}.}.

In Section \ref{sec:symmetrybreaking}, we construct a catalog of
explicit minimal models. Since $\gD$ needs to include a $U(1)$
factor for kinetic mixing with SM hypercharge, we take $\gD =
SU(2) \times U(1)$. Our models differ only in their dark Higgs
sectors, which are constructed to break $\gD$ completely and
induce all the necessary couplings between the different states of
the DM multiplet.

In Section \ref{sec:masssplitting}, we discuss the mass splittings
between the dark matter states. In order to obtain the small mass
splittings needed for XDM and iDM, we consider DM that is a
doublet or a triplet under $SU(2)_{\rm dark}$. The splittings may
be generated radiatively from dark gauge boson loops. Another
possibility is to generate them through higher-dimensional
couplings between the dark matter and a single dark Higgs.

In Section \ref{sec:SUSY}, we consider the addition of SUSY to the
dark sector.  We observe that the minimal assumption of kinetic
mixing between dark sector and SM hypercharge generates an
effective FI term in the dark sector that is naturally of the
desired scale, $\mathcal{O}($GeV$)$.  This term can break SUSY, or even more interestingly can generate a super-Higgs
mechanism that leaves a supersymmetric dark sector with a $\sim 1$
GeV gap.  Both of these scenarios typically result in light
fermions that may have an influence on collider physics.  We
emphasize that this is a {\it leading} contribution which must be
included in any SUSY scenario that includes kinetic mixing.
Furthermore, within this scheme the DM can easily be a SM singlet,
and so DM annihilations do not produce SM $W^\pm$ bosons that
would dangerously decay to anti-protons that have not been
observed by PAMELA\@.  We also investigate the gauge mediation
scenario originally proposed in Ref.\ \cite{ArkaniHamed:2008qp}
where DM is charged under the SM gauge group. An additional
complication we address arises because SUSY restricts the form of
the scalar potential which is responsible for breaking $\gD$
completely. We provide several examples to overcome this
difficulty.

In Section \ref{sec:benchmarks}, we present several benchmark
models for the dark Higgs sector. The resulting spectra of light
vector bosons and scalars are explicitly computed and the relevant
couplings are discussed.

In Section \ref{sec:collider}, we investigate the collider
signatures of these models\footnote{We leave any precise matching
to astrophysical observations for future work.}. The kinetic mixing
is the essential gateway to produce and observe dark sector
states. Dark gauge bosons can be produced in processes analogous
to prompt photon production in the SM\@. They can also be produced
through rare Z decays. The dark sector states themselves
dominantly decay into multiple $e^{\pm}$ and $\mu^{\pm}$, which
are highly collimated and dubbed ``lepton jets''
\cite{ArkaniHamed:2008qp}. Due to the non-abelian structure of
$G_{\rm dark}$, these lepton jets typically contain more than two
leptons each. We discuss the observability of such signals at the
LHC and at the Tevatron. We find that the cascade decays in the
dark sector may result in displaced vertices or possible
correlations between the $\Ptmiss$ and the lepton jet. Several
such displaced vertices will produce uncorrelated impact
parameters of decay products.  In the case where the dark sector
is supersymmetric, then it may be possible to detect direct
electroweak gaugino production with enhanced reach both at the
Tevatron and at the LHC\@. As a bonus, we find that we can exploit
these cascades to perform absolute mass measurements of MSSM
gauginos. Section \ref{sec:conclusions} contains our conclusions.

Finally, let us briefly comment on the notational conventions used
in this paper.  In general, symbols referring to elements of the
SM will be capitalized---so for example the SM hypercharge gauge
coupling, gauge field, and field strength will be denoted by
$g_Y$, $B_\mu$ and $B_{\mu\nu}$.  In contrast, lowercase symbols
will refer to elements of the dark sector, so the dark sector
gauge coupling, gauge field, and field strength will be denoted by
$g_y$, $b_\mu$ and $b_{\mu\nu}$.  We  will use $h$, or $h'$ to denote dark Higgses. We denote the dark matter states
by $\Psi$ and we denote the SM and dark photon by $\gamma$ and
$\gamma^\prime$, respectively.


\section{The Dark Sector and Symmetry Breaking}
\label{sec:symmetrybreaking}
 Let us begin by discussing the basic structure of the dark sector models that we will consider in this
paper. We take the DM to be the lightest (and stable) component of
some multiplet of the non-abelian group $G_{\rm dark}$.  As we
will discuss in Section \ref{sec:masssplitting}, such a multiplet
is necessary if we wish to explain the INTEGRAL and/or DAMA
signals along the lines of the XDM and iDM proposals of \cite{Finkbeiner:2007kk,TuckerSmith:2001hy,Chang:2008gd}.

Furthermore, we follow the proposal of \cite{ArkaniHamed:2008qn} in which the SM is
coupled to the dark sector via a kinetic mixing term between SM
hypercharge and a dark sector $U(1)$ gauge field:
\begin{eqnarray}
\label{eq:kmix}
\mathcal{L}_{\rm gauge} &=&
-\frac{1}{4}B_{\mu\nu} B^{\mu\nu} -\frac{1}{4} b_{\mu\nu}
b^{\mu\nu} +\frac{\epsilon}{2}  B_{\mu\nu}b^{\mu\nu}
\end{eqnarray}
where $B_{\mu\nu}$ and $b_{\mu\nu}$ are the SM and dark sector
hypercharge field strengths, respectively. Because this marginal
operator preserves all of the symmetries of the SM, it is
relatively unconstrained phenomenologically.  For a detailed
analysis of kinetic mixing and the couplings it induces between SM
and dark sector fields, see appendix \ref{app:KinMixing}.

Since $G_{\rm dark}$ must contain a $U(1)$ factor\footnote{It is
actually  possible to achieve mixing without an abelian factor
through an S parameter type operator
$\mathrm{Tr}\left[\Phi w_{\mu\nu}\right]B^{\mu\nu}$, where $\Phi$
is some operator that transforms as an adjoint of the non-abelian
group. In this paper we keep the abelian factor in order to investigate the collider signatures of the more general gauge group structure and ignore the existence of such operators. That is certainly justified in the case  where no fundamental adjoints are present and the contribution is subleading.}, the minimal choice is of course $\gD = SU(2) \times
U(1)$. Furthermore, if $\gD$ is broken completely at a scale of
$\sim$ GeV, then the resulting mass gap will relieve constraints
from BBN on the number of relativistic degrees of freedom.  However,
in order to fully break charge, it is necessary to appropriately
engineer a dark Higgs sector. As we shall see shortly, these
scalars must also break a custodial $SU(2)$ in order to be
phenomenologically viable. The necessity of breaking these
symmetries demands a fairly elaborate dark Higgs sector.

First, let us consider the issue of charge breaking.  Even for the
simplest two Higgs doublet model, the criterion for charge
breaking is quite complicated \cite{Ginzburg:2007jn}, for theories with more
exotic Higgs representations, the space of charge breaking vacua
is not even known.  In appendix \ref{app:CBM}, we present a
straightforward method for deriving necessary conditions for
charge breaking in two higgs doublet sectors, which we applied in
order to obtain viable dark sector benchmark models.

Now let us explain the problem of the custodial symmetry. In the
spirit of \cite{ArkaniHamed:2008qn}, we will
assume that the DM is a multiplet of  $G_{\rm dark}$ whose
components are split in mass. The resulting excited and ground
states have transitions mediated by dark gauge bosons that need to
couple to the SM electric current if they are to realize the XDM
and/or iDM scenarios (see Section \ref{sec:masssplitting}).
However, this mixing can be forbidden by a custodial symmetry of
the dark Higgs sector. To see why this is the case, consider a
model of two scalar doublets. We define $w_{a\mu}$ and $b_\mu$ to
be the gauge bosons of $\gD$, where $b_\mu$ is the abelian field
which mixes with the SM hypercharge, $B_\mu$. Assuming arbitrary
vevs for the scalars, $\gD$ is broken, and in the
$\{w_1,w_2,w_3,b\}$ basis, the gauge boson mass matrix takes the
form
\begin{eqnarray}
\label{eqn:gbMassMat}
M^2_{\rm dark \ gauge } &=& \left(
\begin{array}{cccc} m_w^2 & 0 & 0 & \Delta_1 \\
0 & m_w^2 & 0 & \Delta_2 \\
0 & 0 & m_w^2 & \Delta_3 \\
\Delta_1 & \Delta_2 & \Delta_3 & m_b^2
\end{array} \right)
\end{eqnarray}
As a consequence of the custodial symmetry  present in any theory
of only scalar doublets, the diagonal entries $w_i$ are all equal.
Applying a custodial $SU(2)$ transformation, we can rotate the
components $\Delta_i$ completely into the $w_3$ direction. This
yields a mass matrix which has a manifest $U(1)$ symmetry that
acts as a phase rotation on $w_\pm = w_1 \pm i w_2$ (note that the
gauged ``electromagnetism'' can still be broken while preserving
this $U(1)$). Under this $U(1)$ the components of the DM multiplet
have distinct charges---consequently the gauge bosons that mediate
transitions among these states must also be charged, so they can
only be the $w_\pm$.  However, $w_\pm$ have no components in the
$b$ direction, so they do not kinetically mix with SM hypercharge
and thus cannot decay to SM particles.

Because the custodial symmetry is broken explicitly by the dark
hypercharge, the couplings to the SM  that are excluded at
tree-level by this symmetry will be generated at one loop. Indeed,
this may actually be desirable,
 since it generates an effective coupling for the DAMA transition that
is suppressed beyond the  $\epsilon^2$ from the kinetic mixing.
Another possibility, considered below, is to include additional
Higgses that break custodial symmetry at tree-level.

In the case of SUSY models, we will be forced to significantly
enlarge the Higgs sector.  This is because many of the
difficulties that arise in the non-SUSY case are exacerbated with
the additional constraints imposed by SUSY\@.  Moreover, in SUSY,
all scalars are complex, which forces us to promote real Higgs
triplets to complex Higgs triplets.  This, along with the
constraint of anomaly cancellation implies somewhat of a
proliferation of Higgses in these theories.

In what follows, we enumerate several types of scalar sectors that
break dark charge as well as custodial symmetry. We focus on
models with the intention of later extending them with
supersymmetry.

\subsection{Doublet Models} \label{subsect:2D}

A theory of one Higgs doublet is incapable of breaking charge, so
we consider two doublets
$h_1$ and $h_2$ with quantum numbers ${\bf 2}_{-1/2}$ and ${\bf
2}_{1/2}$ under $G_{\rm dark} = SU(2)\times U(1)$. A general
renormalizable scalar potential that breaks charge is given
by\footnote{This is not the most general renormalizable scalar
potential possible. One can add three more terms,
$|h_1|^2|h_2|^2$, $|h_1|^2h_1^T \epsilon h_2$, and $|h_2|^2h_1^T
\epsilon h_2$ which are consistent with all the symmetries.
However, these simply complicate the potential and are not
required for breaking charge. A more general analysis of the vacuum structure can be found in Ref.~\cite{Ginzburg:2007jn}},
\begin{eqnarray}
\label{eqn:2HDPot}
V(h_1,h_2) &=& \frac{\lambda_1}{2}\left(|h_1|^2
- |v_1|^2 \right)^2 + \frac{\lambda_2}{2}\left(|h_2|^2 - |v_2|^2
\right)^2 \\\nonumber &+& \lambda_4\left| h_1^T \epsilon h_2 - v_1
v_2 \cos\alpha \right|^2 + \lambda_3\left(|h_1|^2 - |v_1|^2
\right) \left(|h_2|^2 - |v_2|^2 \right)
\end{eqnarray}
with,
\begin{equation}
\label{eqn:quarticConds}
\lambda_1 > 0,\quad \lambda_2>0,\quad \sqrt{\lambda_1\lambda_2} +\lambda_3 > 0,  \quad \quad \lambda_4 > 0
\end{equation}
$v_2$ is complex and charge is broken when $0 < |\cos\alpha|< 1$.

In the MSSM the conditions of Eq. (\ref{eqn:quarticConds}) are
violated at tree level. From the D-term contributions to the
scalar potential we have $\lambda_1 = \lambda_2 = -\lambda_3 =
(g^2+\gps)/8$. The inequality is saturated and the potential in Eq. (\ref{eqn:2HDPot}) degenerates and contains a flat direction. To avoid such flat directions in the MSSM one must usually evoke a condition on the quadratic terms in the potential. Such potentials cannot be placed in the form of Eq. (\ref{eqn:2HDPot}) and charge is not broken. Therefore the usual supersymmetric two doublets model will not suffice and we need additional contributions to the scalar potential
in order to satisfy the condition, Eq. (\ref{eqn:quarticConds}).

In addition, since this model has a custodial symmetry, it fails
to have proper mixings between the gauge bosons.  Nonetheless,
since the custodial symmetry is broken by dark hypercharge, the
gauge boson mixing receives one-loop radiative corrections that
break the custodial symmetry.  From this point of view there is
also no reason not to include higher dimension custodial violating
operators that can be generated if heavy (triplet) states have
been integrated out.  In fact, we include such irrelevant
operators in the benchmark model of Section \ref{subsubsec:nonSUSY1}.

Here we also note the presence of an unfortunate $\mathbb{Z}_2$
symmetry that is present in the $\tan \beta = 1$ limit.  This
symmetry needs to be broken since it prevents two of the dark
gauge bosons from coupling to SM electric charge (see appendix
\ref{app:Z2}).

\subsection{Doublet/Triplet Models} \label{subsect:1D1T2S}

An obvious way to break custodial $SU(2)$ at tree-level is to
augment the two doublet model with a light triplet of $SU(2)$. For
instance, consider a model of one doublet, $h$, and one real
triplet, $\Phi$, with dark quantum numbers ${\bf 2}_{1/2}$ and
${\bf 3}_{0}$, respectively. In order to realize a charge breaking
angle between the doublet and triplet, we include the following
two operators: $h^\dagger \Phi h$ and $
  h^T \epsilon \Phi h$.  Since the latter has nonzero
hypercharge, we must multiply it by a new hypercharged singlet,
$S$, in order to include it in the potential:
\begin{eqnarray}
\label{eqn:1D1T2SP} V(h,\Phi,S) &=& \frac{\lambda_h}{2}\left(|h|^2
- |v_h|^2 \right)^2 +
\frac{\lambda_\Phi}{2}\left(\mathrm{Tr}\left[\Phi \Phi \right] -
|v_\Phi|^2 \right)^2 + \frac{\lambda_S}{2}\left(\left| S
\right|^2- |v_S|^2 \right)^2 \\\nonumber &+& c_1 h^\dagger \Phi h
+ (c_2 S h^T \epsilon \Phi h + \textrm{h.c.})
\end{eqnarray}
Alternatively, we might consider a model with two doublets and one
triplet.  This is more natural if we wish to eventually include
SUSY\@.  The scalar potential takes the form:
\begin{eqnarray} \label{eqn:2D1TP}
V(h_1,h_2,\Phi) &=& V(h_1,h_2) +
\frac{\lambda_\Phi}{2}\left(\mathrm{Tr}\left[\Phi \Phi \right] -
|v_\Phi|^2 \right)^2 \\\nonumber &+& c_1 h_1^\dagger \Phi h_1 +
c_2 h_2^\dagger \Phi h_2 + \left(c_3 h_1^T \epsilon \Phi h_2 +
\mathrm{h.c.} \right)
\end{eqnarray}
where $V(h_1,h_2)$ is the contribution from doublets alone defined
in eq \ref{eqn:2HDPot}.

We can impose an additional $\mathbb{Z}_2$ symmetry $\Phi
\rightarrow -\Phi$, that forbids tree-level couplings between the
triplet and doublets: $c_1=c_2=c_3=0$.  This enhanced global
symmetry implies the existence of two pseudo-Goldstone bosons
which obtain masses at one-loop $\sim 10$ MeV\@. These
pseudo-Goldstone bosons will be produced at the bottom of dark
sector cascades.  They decay into leptons through either two off-shell dark gauge bosons or at one-loop (see Fig.\ \ref{fig:pseudo-decay} and the discussion in Section \ref{subsubsec:lepjetstruc}).  Either way, the long lifetime causes the pseudo-Goldstone boson to escape the detector at colliders. Since those are pseudo-scalars they will not contribute to the Sommerfeld enhancement of DM annihilations in the early universe and their mass is therefore not bounded by the limits derived in Ref. \cite{Kamionkowski:2008gj}.


\section{Dark Matter Mass Splitting}
\label{sec:masssplitting}

The authors of \cite{ArkaniHamed:2008qn} observed that a DM
multiplet of some non-abelian $G_{\rm dark}$, given appropriate
mass splittings, can in principle realize the XDM explanation of
INTEGRAL \cite{Finkbeiner:2007kk} and also the iDM mechanism for
reconciling the DAMA annual modulation with the null result of
other direct detection experiments
\cite{TuckerSmith:2001hy,Chang:2008gd}.  In this section we
briefly review these proposals, and discuss concrete ways of
generating the appropriate mass splittings within concrete
theories.

The INTEGRAL collaboration has provided an extremely refined measurement of the $511\keV$ line of positronium annihilation coming from the galactic center. In the XDM scenario, WIMPs in the galactic center scatter into an excited state, lying $\sim 1\MeV$ above the ground state. The excited state then de-excites into $e^+e^-$ which provides the excess positrons needed. In terms of model-building we need a splitting of  $\sim1\MeV$ between two states in the DM multiplet. Transitions between these two states are mediated by a dark gauge boson with some component of the dark hypercharge (which in
turn couples to SM leptons).

In contrast, DAMA is a direct detection experiment which seeks to
measure the scattering of galactic WIMPS off of NaI(Tl).  Assuming
a standard WIMP with elastic scattering, several other experiment
such as CDMS\cite{Akerib:2005kh,Ahmed:2008eu}, XENON
\cite{Angle:2007uj}, and ZEPLIN \cite{Lebedenko:2008gb} exclude
DAMA's measured annual modulation by many orders of magnitude. The
iDM proposal reconciles these experiments by proposing that the
WIMP can only scatter off of nuclei through an inelastic process
by which the the DM is converted into a slightly excited state.
Since the WIMP kinetic energy is fixed and the threshold for the
inelastic transition is dependent on the atomic number of the
nuclei, this iDM scenario can simultaneously predict a null result
at CDMS and a positive result at DAMA\footnote{XENON and ZEPLIN,
which both use Xe as a target should be able to exclude the iDM
scenario, but at the moment these experiments are background
limited \cite{Chang:2008gd}}.  Considering fermionic DM, this
scenario can be accommodated by including a mass splitting of
around $\sim$ 100-150 keV \cite{TuckerSmith:2001hy,Chang:2008gd}
between the lightest two Majorana states of the fermion.  The
bottom line for model building is that to evade CDMS and
CRESST\cite{Angloher:2008jj} bounds, the DM must be split from the
next heaviest Majorana state by at at least $100\keV$.

Before we consider mechanisms for generating the required splittings, we must ascertain that there is no elastic scattering which would have been seen in direct detection experiments.  One possibility is to begin with Majorana dark matter in a real representation of the dark gauge symmetry.  Gauge bosons then couple different states of the multiplet and radiative corrections, to be discussed in Section \ref{subsec:radiativesplitting}, can split the masses of these states.  But if the dark matter begins in a complex representation, for example if it has dark or SM U(1) charge, then it must be Dirac-like at high-energies.  Then the model is already excluded by direct detection experiments since the elastic scattering of Dirac-like dark matter is not sufficiently suppressed unless $\epsilon \lesssim10^{-6}$.  However, it is possible to split the masses of the Majorana components of the Dirac fermions by using the same scalar sector that is responsible for breaking dark gauge symmetry.  For instance, if we imagine that $\phi$ is some scalar singlet
whose vev breaks global fermion number and $U(1)_{y}$, we can add
a term such as,
\begin{equation}
\mathcal{L}_{\rm Majorana} = \phi \Psi \Psi + \phi^* \Psi^c \Psi^c
\end{equation}
where $\Psi$ and $\Psi^c$ are the Weyl components of some DM multiplet. If $\phi$
develops a vev of order $\sim\GeV$, it will generate a Majorana
mass splitting that forbids any elastic scattering and evades direct detection
bounds.

Another possibility is to use a higher dimensional operator with a dark sector doublet, $h$,
\begin{equation}
\mathcal{L}_{\rm Majorana} = \frac{1}{M_X} h \Psi \Psi h,
\end{equation}
where $M_X \sim \TeV$.  In this case, the Majorana splitting is of
order $\sim \MeV$ which again kinematically forbids elastic
scattering.  If DM is charged under the SM as part of a
$\mathbf{5} + \mathbf{\bar 5}$ multiplet, then a dimension 6
operator is required to contract both dark hypercharge and SM
quantum numbers.  For example, we can use the operator $1/M_X^2 \,
H \Psi \Psi H \phi$, where $H$ is the SM Higgs, $\phi$ is a
singlet that soaks up $\Psi$'s dark hypercharge and gets a vev at
$\sim1\unit{GeV}$, and  $M_X \sim \unit{TeV}$.  This leads to
Majorana splitting of order $\sim$ GeV, which forbids elastic
scattering.

Any of the possibilities mentioned above can be employed to evade direct detection from CDMS\@.  In the next
subsection, we consider two possible way for generating the appropriate $\sim1\MeV$ and $\sim 100\keV$ mass splittings necessary for XDM and iDM\@.

\subsection{Radiative Splitting}
\label{subsec:radiativesplitting}

As is well-known \cite{Thomas:1998wy}, spontaneous symmetry
breaking of a non-abelian gauge group  generates radiative mass
splittings within a multiplet of the symmetry. We
take the DM multiplet to have mass $\sim 500-800$ GeV, and to be charged under
the dark $SU(2)\times U(1)$.  As discussed in section
\ref{sec:symmetrybreaking}, realistic dark sectors must break
charge and custodial $SU(2)$, but to develop some intuition about
the radiative mass splittings, we will begin by considering the
limit where these symmetries are preserved.  In this limit the
mass splittings among the multiplet take a particularly simple
form,
\begin{eqnarray}
\label{eqn:radmasssplit} \Delta m_{ij} &=& \frac{\alpha^{\rm
dark}}{2}  (q_i^2 - q_j^2)M_z \\\nonumber &-&\frac{\alpha_2^{\rm
dark}}{2}\left((T^3_i)^2-(T^3_j)^2\right)\left(M_z-M_w\right),
\end{eqnarray}
where we define $\alpha_2^{\rm dark}$, and $\alpha^{\rm dark}$ as
usual with respect to $SU(2)\times U(1)$ couplings. The charges
are $q_i = T^3_i+Y$ and $T^3_i$ is the $i^{th}$ eigenvalue of the
third $SU(2)$ generator. In the more general limit where charge
and custodial symmetry are broken, one must use the appropriate
vector boson mass eigenstates and their couplings to the fermions
in order to compute the mass correction (Eqs.
\ref{eqn:massCorrLoops}). This is a straightforward computation,
however, in general it does not yield a simple analytic result.
Nevertheless, it is clear that there are two factors which control
the mass splitting: first, the differences between masses of the
vector bosons; and second, the couplings of the different members
of the representation to the vector bosons.

\begin {figure}
\center
\includegraphics[width=0.6\textwidth]{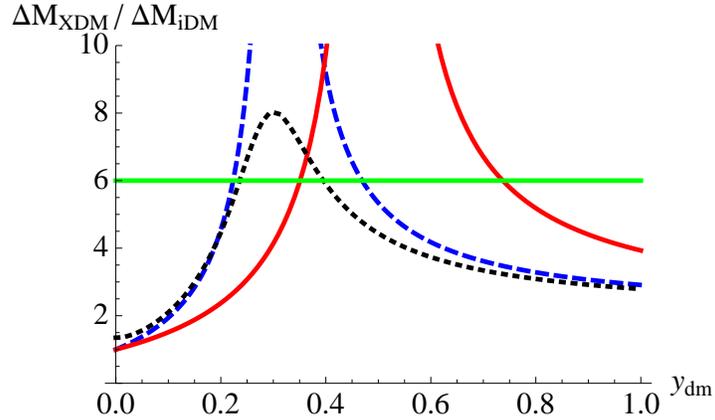}
\caption{\label{fig:2D1T_1Dplot} The ratio of the XDM splitting to
the iDM splitting as a function of triplet dark matter $U(1)_y$
hypercharge.  The green horizontal line indicates the minimum
ratio for simultaneously achieving both splittings.  Red (line) is
an example of two Higgs doublets with charge preserved, blue
(dashed) represents two Higgs doublets with charge broken, and
black (dots) adds a Higgs triplet to the previous case.   For this
example, the gauge couplings are $g=0.97$ and $g_y=0.26$, and in
terms of Eq.~\ref{eqn:2HDPot} we have for all three models
$v_1=0.9 \unit{GeV}$, $v_2=1.1 \unit{GeV}$ and $\lambda_{1,2,3,4}
= 1$.  Red (line) and black (dots) add charge breaking with $\cos
\alpha = 0.75$, and for black (dots), in terms of
Eq.~\ref{eqn:2D1TP}, $\lambda_\Phi=1$, $v_\Phi = 1 \unit{GeV}$ and
the triplet is decoupled from the doublets at tree-level by
imposing the discrete symmetry: $\Phi \rightarrow - \Phi$. }
\end {figure}

\begin {figure}
\center
\includegraphics[scale=0.6]{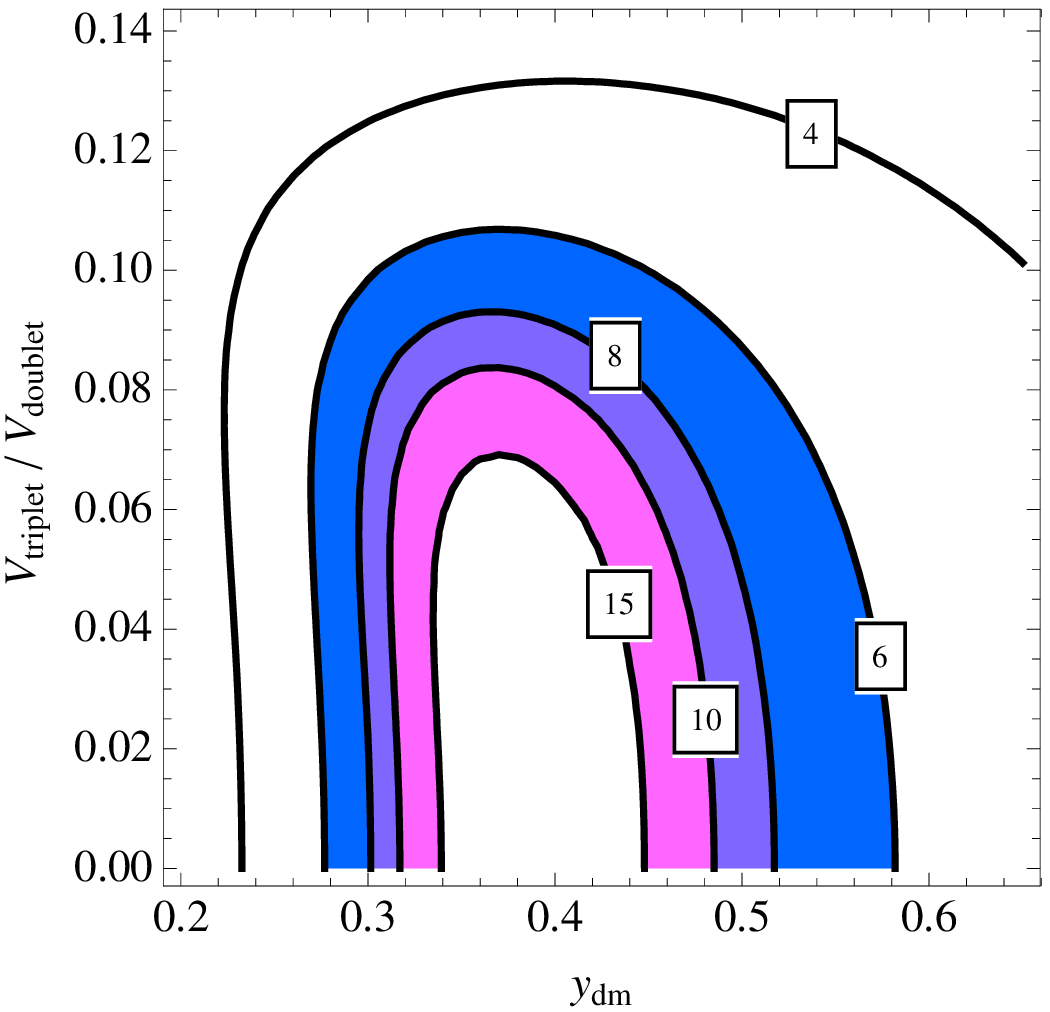}
\hspace{1cm}
\includegraphics[scale=0.66]{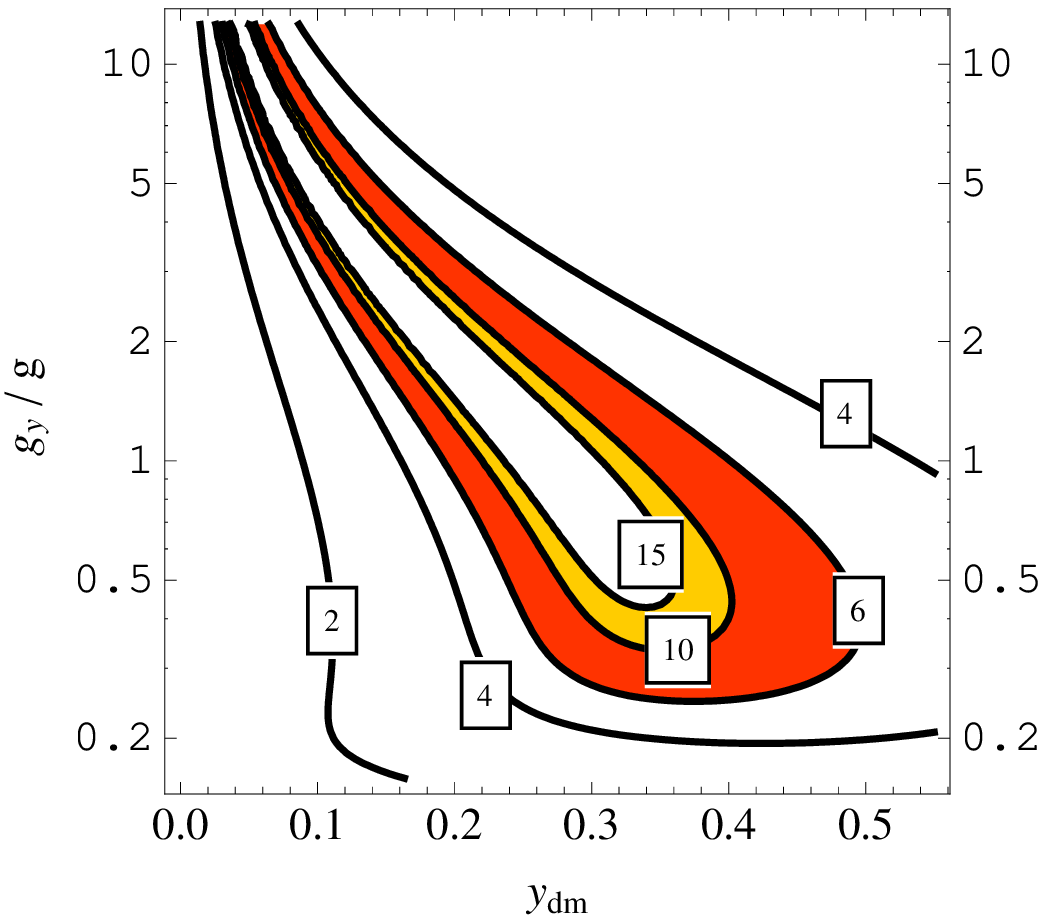}
\caption{\label{fig:2D1T_Conts}
Two contour plots of the ratio of the XDM splitting to the iDM
splitting for triplet dark matter with two Higgs doublets and one Higgs
triplet.   The shaded regions represent splitting ratios where XDM and iDM can be achieved simultaneously.  In both plots, the horizontal axis is the dark matter $U(1)_y$
hypercharge.  The vertical axis of the left plot represents the ratio of the triplet to doublet VEVs, $v_\Phi / v$, where $v^2=v_u^2+v_d^2$ and $\left< \Phi\right> = v_\Phi T_3$.  The vertical axis of the right plot represents the ratio of dark hypercharge and $SU(2)$ couplings, $g' / g$.  For both plots, the triplet is decoupled from the doublets at tree-level by imposing the discrete symmetry: $\Phi \rightarrow - \Phi$, and in terms of Eq.~\ref{eqn:2HDPot} we have $v_1=0.9 \unit{GeV}$, $v_2=1.1 \unit{GeV}$, $\cos \alpha = 0.9$ and $\lambda_{1,2,3,4} = 1$.  For the left plot, the gauge couplings are $g=0.97$ and $g_y=0.26$.  For the right plot, we have also chosen, in terms of Eq.~\ref{eqn:2D1TP}, $\lambda_\Phi$ = 1 and $v_\Phi = 1 \unit{GeV}$.
}
\end {figure}

As a simple example with all the required splittings and couplings
we can consider a triplet with hypercharge $y = 1/2 -\delta$. We
generate both large, $\Delta M \sim \alpha^{\rm dark} M_z$ and
small $\Delta m \sim \delta \alpha^{\rm dark} M_z$ splittings.
The correct couplings to account for the XDM and iDM scenarios are
induced when charge and custodial breaking corrections are
included.  A realistic model certainly need not be based on such
odd charge assignments, however, this example serves to illustrate
how straightforward it is to obtain the correct splittings and
couplings.  In Figs. \ref{fig:2D1T_1Dplot} and
\ref{fig:2D1T_Conts}, we consider some of the more general models
of Section \ref{sec:symmetrybreaking}, which include charge and
custodial symmetry breaking, and we plot the exact ratio of the
two splittings relevant to XDM and iDM as a function of the
parameters. The corrections induced in supersymmetric models are
discussed in appendix \ref{app:SUSYcont}.

\subsection{Mass splitting from higher dimensional operator}
\label{subsec:hdoperatorsplitting}

It is also possible to generate the
INTEGRAL and DAMA mass splittings from higher dimension operators
alone. The key observation is that $\delta m \sim \Lambda_{\rm
dark}^2/M_X \sim $ MeV, which is of the desired range.

As an example, we consider two Weyl fermions $\Psi, \Psi^c$ which
are $\textbf{2}_{1/2}$ and $\bar{\textbf{2}}_{-1/2}$ under $G_{\rm
dark}$. It is possible to achieve all the required splittings and
transitions with a single scalar doublet,
\begin{equation}
\mathcal{L} \supset M_{\Psi} \Psi \Psi^c + \frac{\lambda_1}{M_X}
\Psi h \Psi h +  \frac{\lambda_2}{M_X} \Psi^c h^c \Psi^c h^c
+  \frac{\lambda_3}{M_X} \Psi^c h^c \Psi h + h.c.
\end{equation}
with $\Psi = (\psi_\nu,\psi_e)$ and  $\Psi h \equiv \psi_i
\epsilon_{ij} h_j$ and $h^c_i = \epsilon_{ij} h^*_j$. Once the
scalar doublet gets a vev, $\langle h\rangle = (0,v)$, the
``neutrino'' components of $\Psi$ and $\Psi^c$ mix through the
following matrix,
\begin{equation}
M= \left(\begin{array}{cc}\lambda_1 \bar{v} & M_{\Psi}+\lambda_3
  \bar{v} \\ M_{\Psi}+\lambda_3 \bar{v} & \lambda_2 \bar{v}  \end{array}
\right),
\end{equation}
where $\bar{v}= v^2/M_X \sim $ MeV\@.
In the limit where $\lambda_1= \lambda_2 =0$ the states
are maximally mixed, $\psi^{\pm} = \left( \psi_\nu\pm\psi_\nu^c\right)/\sqrt{2}$, and form a Dirac pair of mass $M_{\Psi}+\lambda_3 \bar{v}$ which provides the XDM splitting when compared with the $\psi_e,\psi_e^c$ states of mass $M_\Psi$.

\begin{figure}
\begin{center}
\includegraphics[scale=0.7]{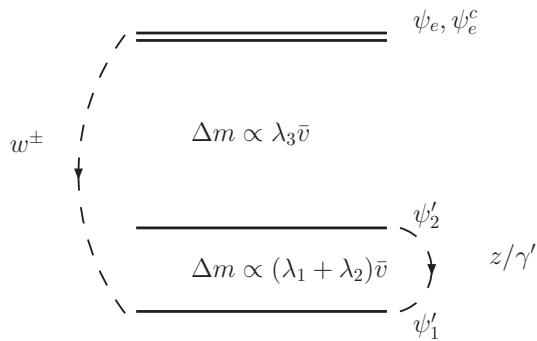}
\end{center}
\caption{The resulting spectrum for a Dirac doublet with majoron coupling.}
\label{fig:YukawaSplit}
\end{figure}

With non-zero $\lambda_1$ and $\lambda_2$ we have,
\begin{eqnarray}
\psi_1' &=& \cos\theta \psi^+ + \sin\theta \psi^- \quad\quad\quad m_1 =
M_{\Psi}+\frac{\bar{v}}{2}(2\lambda_3 + \lambda_1+\lambda_2) \\\nonumber
\psi_2' &=& -\sin\theta \psi^+ + \cos\theta \psi^- \quad\quad\quad m_2 =
-M_{\Psi}-\frac{\bar{v}}{2}(2\lambda_3 - \lambda_1 - \lambda_2)
\end{eqnarray}
with $\sin\theta \approx (\lambda_1-\lambda_2)\bar{v}/4 M_\Psi$. The mass difference
between the two states is $|\Delta m_{12} | = (\lambda_1+\lambda_2)
\bar{v}$. So, by tuning $\lambda_1$ against $\lambda_2$ we can achieve
$\Delta m_{12} \sim 0.1 \MeV = 100 \keV$ as required by iDM\@. The coupling of the mass
eigenstates to the dark gauge boson is given by,
\begin{eqnarray}
g_y \bar{\Psi}\slashed{b}\Psi - g_y
\bar{\Psi}^c\slashed{b}\Psi^c &=& g_y
\sin\theta \cos\theta \, (\bar{\psi}_2'\slashed{b}\psi_2' -
\bar{\psi}_1'\slashed{b}\psi_1') \nonumber \\ &-& g_y \cos^2\theta \,
\bar{\psi}_1'\slashed{b}\psi_2' + h.c.
\end{eqnarray}
In this case the ratio of the elastic to inelastic coupling is
approximately~$\sin\theta = (\lambda_1-\lambda_2)(\bar{v}/M_{\Psi}) \sim 10^{-7}$, which is
sufficiently suppressed. The spectrum relevant for this case is shown in Fig.~\ref{fig:YukawaSplit}.

\section {Generation of the Dark Sector Mass Scale}
\label{sec:SUSY}

As noted in \cite{ArkaniHamed:2008qp}, a particularly nice feature
of a SUSY dark sector is that the GeV scale is naturally generated
by gauge mediated SUSY breaking from the SM\@. In this section, we
elaborate on this scenario in detail. Furthermore, we propose an
even more minimal alternative in which ``kinetic mixing
mediation'' breaks SUSY or induces a super-Higgs
mechanism at a scale of several GeV in the dark sector.  As we
will discuss, these theories typically have light fermions which
affect the collider physics.

For gauge mediation, dark matter itself can act as the messenger
if we take it to be charged under the SM as part of a $\mathbf{5}
+ \mathbf{\bar 5}$ multiplet. Dark matter annihilations then also
produce SM electroweak gauge bosons, resulting in hadronic
channels.  But since the GeV scale can be generated by kinetic
mixing mediation alone, there is no need to charge dark matter
under the SM\@.

Although we focus on kinetic mixing mediation and gauge mediation
for the rest of this section, and when we construct benchmarks in
Section \ref{sec:benchmarks}, there are other ways to break SUSY
in the dark sector.  We would like to stress that the rest of our
paper, in particular the model-independent discussion of collider
signatures in Section \ref{sec:collider}, does not depend on how
SUSY is broken in the dark sector. One alternative is that there
is high-scale gauge mediation and a GeV scale gravitino
\cite{ArkaniHamed:2008qp}. Then SUSY is broken in the dark sector
at the GeV scale by a ``Planck slop.''  Another possibility is
that the dark matter mass is related to the mechanism that sets
the MSSM $\mu$ parameter, for example due to a superpotential of
the form: $\lambda \, S H_u H_d \; + \; \lambda' \, S \Psi
\Psi^c$.  A vev for $F_S$ is communicated to the dark sector
through gauge mediation with dark matter as messengers.

\subsection {SUSY Breaking from Kinetic Mixing} \label{subsect:KMBreak}

In \cite{Dienes:1996zr} it was observed that mixing of gauge boson kinetic
terms will induce mixed D-term contributions to the action that
can communicate SUSY breaking between two sectors that are
otherwise decoupled.\footnote{Strictly speaking, this is a form of
gauge mediation according to the definition of Ref.~\cite{Meade:2008wd}}  There the authors noted new, possibly dangerous
contributions to SUSY breaking to the MSSM from this effect.  In
this section, we use this effect to our advantage in order to
mediate SUSY breaking from the SM to the dark sector.  We should
note that while we can choose to make this the dominant mechanism
for breaking SUSY in the dark sector, it is always present at the
GeV scale.\footnote{Kinetic mixing mediation is neglected in some recent $U(1)$ dark sector papers, for instance Ref.\ \cite{Zurek:2008qg} focuses on a form of mediation that is sub-leading in $\epsilon$.}

As was originally proposed in \cite{ArkaniHamed:2008qn}, we have been assuming that
the dark sector and the SM are coupled via a marginal gauge
kinetic mixing between the dark hypercharge, $U(1)_y$ and the SM
hypercharge, $U(1)_Y$.  If both $U(1)$'s are fundamental, then the kinetic mixing is a UV boundary condition, sensitive to physics at the highest scales.  Instead, if either $U(1)$ is embedded in a GUT, then the kinetic mixing is only induced below the scale of GUT breaking by integrating out fields charged under both $U(1)$'s.  In this case we can estimate its size.   In particular, heavy fields charged under both the SM and the dark sector will induce a gauge kinetic mixing:
\be \label{eqn:KMsuperfield} \cL_{\mathrm{gauge}} &=& \frac{1}{4}
\int d^2\theta \left( W_{Y}
W_{Y} + W_y W_y - 2 \epsilon W_{Y} W_y  + \mathrm{h.c.} \right)\\
\epsilon &\sim& -\frac{g_Y g_y}{16\pi^2} \log
\left(\frac{M^2}{M'^2}\right) \ee
where $g_y$ and $g_Y$ are the gauge couplings for the dark and SM
hypercharges, respectively, $M$ and $M'$ are the masses of
components of the heavy particle multiplet. Assuming that these
mass scales are not too separated and that the gauge couplings are
of reasonable size, this gives an estimate of $\epsilon \sim
10^{-3}-10^{-4}$. Interestingly, this not only gives the right
scale to explain the DAMA cross-section, but also generates a
scale of around a GeV in the dark sector.  Eq.
(\ref{eqn:KMsuperfield}), along with the Kahler potential, implies
a D-term potential:
\begin {equation} \label{eqn:KMpotential}
V_{\mathrm{gauge}} = \frac{1}{2}D_Y^2 +
\frac{1}{2}D_{y}^2-\epsilon D_Y D_y + g_Y D_Y \sum_i Q_i \left|
H_i \right|^2 + g_y D_y \sum_i q_i \left| h_i \right|^2
\end {equation}
where $H_i$ and $h_i$ denote the SM and dark sector Higgs,
respectively, and $Q_i$ and $q_i$ denote their SM and dark sector
hypercharges. Integrating out the SM fields, $H_i$ and $D_Y$,
generates a cross term $\epsilon D_y \langle D_Y\rangle$ in the
low-energy theory.  Thus, in the infrared, this
 induces an effective
Fayet-Iliopolous D-term for $D_y$, for which
\be
\label{eqn:KMFIDterm} V_{\rm gauge} &\supset& \epsilon D_y \langle
D_Y\rangle = \xi D_y
\\
\xi &=& \epsilon \langle D_Y\rangle = \epsilon \frac{g_Y}{2}
\cos{2 \beta} \hspace{2pt} v^2 \ee
where in the last equality we have substituted in for $\left<
D_Y \right>$ from the MSSM\@. For $\epsilon \sim 10^{-3} - 10^{-4}$,
$\xi$ is at the $\unit{GeV}^2$ scale.  Thus, given the minimal
assumption of kinetic mixing and SUSY, we obtain precisely the
right scale to account for PAMELA and ATIC with the Sommerfeld
enhancement.

With an effective FI term at low energies, it is straightforward
to break SUSY in the dark sector.  In particular, a generic
superpotential for the dark Higgses, $h_i$, will break SUSY
because the $F$ and $D$ terms cannot be simultaneously set to
zero.  While this SUSY breaking generates scalar soft masses, it
does not generate soft masses for gauginos.  Moreover, since SUSY
is broken within the dark sector this will typically
introduce a massless dark sector Goldstino\footnote{The gravitino
will eat a linear combination of this field and the Goldstino
associated with SUSY breaking in the MSSM}, assuming the absence
of explicit SUSY breaking operators. So, within this mechanism
there are light fermions. We present a concrete model of this type
in Section \ref{sec:benchmarks}, and mention the possibility of
associated missing energy signals in Section \ref{subsubsec:MET}.

In the opposite extreme, we can take the superpotential to be less
generic, or perhaps even trivial, and so the dark Higgs potential
is dominated by D-terms. Here the dark Higgses simply align to set
the dark hypercharge D-term to zero.  In this limit SUSY is
actually preserved in the dark sector, but a super-Higgs mechanism
will generate a GeV scale dark sector that may still be consistent
with a Sommerfeld enhancement and $\alpha m_z$ mass splittings for
DM\@.   A more minimal superpotential also can imply the existence of light pseudo-Goldstone fields and their superpartners.  Finally, we note that unlike the SUSY breaking case, this super-Higgs scenario will also generate GeV scale gaugino masses.

\subsection {SUSY Breaking From $\mathbf{5} + \mathbf{\bar 5}$ Messengers} \label{subsect:55barMess}

In this section, we elaborate on the gauge mediation proposal of
\cite{ArkaniHamed:2008qp} in which a multiplet of $\mathbf{5} +
\mathbf{\bar 5}$ messengers is charged directly under both the
dark sector and SM, thereby communicating SM SUSY breaking to the
dark sector.  We consider the additional possibility that the
lightest component of the messenger supermultiplet is in fact the
DM\@.

Let us determine the various contributions which set the scale of
masses for the scalar and fermion components of the DM $\mathbf{5}
+ \mathbf{\bar 5}$ multiplet. First, we assume that the fermions
have a SUSY mass, $m_f^{({\bf 2},{\bf 3})}$, that splits the
doublet and triplet components.  Second, in the case where
low-scale gauge mediation explains SUSY breaking in the MSSM, then
the doublet and triplet scalars of the $\mathbf{5} + \mathbf{\bar
5}$ receive identical soft mass contributions to that of the
sleptons and right-handed down squarks of the MSSM\@. We denote this
contribution by $m_s^{({\bf 2},{\bf 3})}\sim 100$ GeV\@. Finally,
the scalar DM can in principle also receive soft mass
contributions from whatever dynamics set its $\mu$ and $B\mu$
terms, which we denote by $B\mu^{({\bf 2},{\bf 3})}$.  Instead of
specifying these dynamics, we will choose a model-independent
parameterization for the DM supermultiplet masses.  The scalar
doublets and triplets of the $\mathbf{5} + \mathbf{\bar 5}$ have a
scalar mass matrix given by:
\be \label{eq:dmmassmat} M_{{\bf 2},{\bf 3}}^2&=& \left(\begin{array}{cc} \left[m_f^{({\bf 2},{\bf 3})}\right]^2 +
\left[m_s^{({\bf 2},{\bf 3})}\right]^2 & B\mu^{({\bf 2},{\bf 3})} \\
B\mu^{({\bf 2},{\bf 3})}  & \left[m_f^{({\bf 2},{\bf 3})}\right]^2 +
\left[m_s^{({\bf 2},{\bf 3})}\right]^2\\
\end{array} \right)
\ee whose eigenvalues, $m_{\pm}^{({\bf 2},{\bf 3})}$, are given by
\begin {equation} \label{eqn:DMmass}
\left(m_{\pm}^{({\bf 2},{\bf 3})}\right)^2  =  \left(m_f^{({\bf 2},{\bf 3})}\right)^2 +
\left(m_s^{({\bf 2},{\bf 3})}\right)^2 \pm B\mu^{({\bf 2},{\bf 3})}.
\end {equation}
The lightest component
of the doublet supermultiplet corresponds to dark matter, and we choose it to have mass $\sim 500 - 800
\unit{GeV}$, which is favored by ATIC\@.

Note that the messenger supertrace of the $\mathbf{5} +
\mathbf{\bar 5}$ mass matrix is non-zero, and proportional to
$m_s^{({\bf 2},{\bf 3})}$. As discussed in \cite{Poppitz:1996xw},
this non-zero supertrace generates a logarithmically UV sensitive
soft mass for the dark sector scalars.  Indeed, since the
messenger supertrace is positive, this implies a negative soft
mass for the dark Higgs:
\begin {equation} \label {eqn:55barMs}
m_h^2 \approx - 8 \left( \frac{\alpha}{4 \pi} \right)^2 \left(2 \left[M_s^{({\bf 2})}\right]^2 +3 \left[M_s^{({\bf 3})}\right]^2 \right)  \log \left( \frac{\Lambda_{UV}^2}{m_f^2} \right) C_a S_q
\end {equation}
where $C_a$ is the dark scalar's quadratic Casimir, $S_Q$ is dark
matter's Dynkin index, and $\Lambda_{UV}$ is set by the messenger
scale of SUSY breaking to the SM\@.  The negative soft mass squared
allows for $G_{\rm dark}$ to break.  This gives us a way to break the symmetries, independent of the effect of RGE running.  It is our assumption that the contributions due to running are suppressed.
For low-scale gauge mediation, $\Lambda_{UV} \sim 30 - 100 \unit{TeV}$, and
because of this logarithmic enhancement and the combined effect of five
messengers, we find that our desired scale of $m_h^2 \sim 1
\unit{GeV}^2$ implies that $m_s^{({\bf 2},{\bf 3})} \sim 50 \unit{GeV}$.  This
indicates a bit of tension numerically because we expect that
$m_s^{({\bf 2},{\bf 3})}$ is set by the SM soft mass scale of hundreds of GeV\@.

Additionally, if we want fermionic DM, then there
is the additional constraint that the fermion is the lightest
component of the dark matter supermultiplet: thus
$(B\mu^{({\bf 2})})^{1/2} < m_s^{({\bf 2})} \ll m_f^{({\bf 2})}$. Since the DM
$B\mu$ contribution breaks the dark sector R-symmetry, the gaugino
soft masses are suppressed if we assume that the triplet component
satisfies the same condition:

\begin {equation} \label {eqn:55barMlambda}
m_\lambda \approx \frac{\alpha}{2} S_q \left( 2
\frac{B\mu^{({\bf 2})}}{m_f^{({\bf 2})}} + 3 \frac{B\mu^{({\bf 3})}}{m_f^{({\bf 3})}} \right)
\end {equation}
This implies light gauginos and the generic prediction is that
fermionic dark matter implies that the lightest dark sector
particle is a mostly-gaugino fermion.  This conclusion can be
avoided by raising $B\mu^{({\bf 3})}$ while maintaining
$(B\mu^{({\bf 2})})^{1/2} \ll m_f^{({\bf 2})}$.

The dark sector Higgses require GeV scale $\mu$ and $B\mu$ terms to help break dark gauge symmetry and lift runaway directions.  These terms can be generated by additional dynamics that communicate SM SUSY breaking to the dark Higgses, in general also resulting in new two-loop contributions to the dark scalar masses. We will assume that these contributions to $m_h^2$ are subdominant to the usual gauge mediation contributions of Eq.~\ref{eqn:55barMs}.  A recent paper identifies a class of general gauge mediation models that satisfy this assumption \cite{Komargodski:2008ax}.

Let us note that while Eqs.~\ref{eqn:55barMs} and \ref{eqn:55barMlambda} are approximations, our benchmark model of gauge mediation in Section \ref{subsubsec:SUSYbench2} employs the full expressions of Ref.~\cite{Poppitz:1996xw}.


\section {Benchmark Models}
\label{sec:benchmarks}

In this section, we present four detailed benchmark dark sector models and their spectra.  The models break dark gauge symmetry and custodial symmetry, generating the dark matter splittings and couplings necessary to explain the astrophysical data, as explained in Section \ref{sec:symmetrybreaking}.  These examples illustrate some of the theoretical issues discussed above, and their spectra and couplings serve as starting points for thinking about the types of cascades that can occur in GeV scale dark sectors.  We begin in Section \ref{subsec:nonSUSYbench} with two non-SUSY models, where the GeV scale is put into the scalar potential by hand.  We then consider two SUSY examples in Section \ref{subsec:SUSYbench}, where the GeV scale is generated radiatively in the dark sector from interactions with the Standard Model.

For each example we consider an $SU(2) \times U(1)_y$ dark sector and triplet dark matter.  We take the Majorana components of the dark matter fermions to be split by enough to avoid direct detection bounds, for example by one of the mechanisms discussed in Section \ref{sec:masssplitting}.  We then calculate the radiative splittings among the triplet, induced by dark symmetry breaking, as in Section \ref{subsec:radiativesplitting}.  We take the ground state to correspond to dark matter, the heaviest excited state to allow for the XDM explanation of INTEGRAL, and the first excited state to allow for the iDM explanation of DAMA\@. We allow complex parameters to carry imaginary parts in order to avoid unbroken CP symmetry in the dark sector which may lead to stable states. This is not necessarily a problem and may actually have additional interesting signatures, but we'd like to keep the spectrum as general as possible for the present discussion. 

\subsection{Non-SUSY Benchmark Models}
\label{subsec:nonSUSYbench}

\subsubsection{Non-SUSY 1: Two Doublets}
\label{subsubsec:nonSUSY1}

\begin {figure}[!h]
\hspace {-1.25cm}
\includegraphics[width=1.2\textwidth]{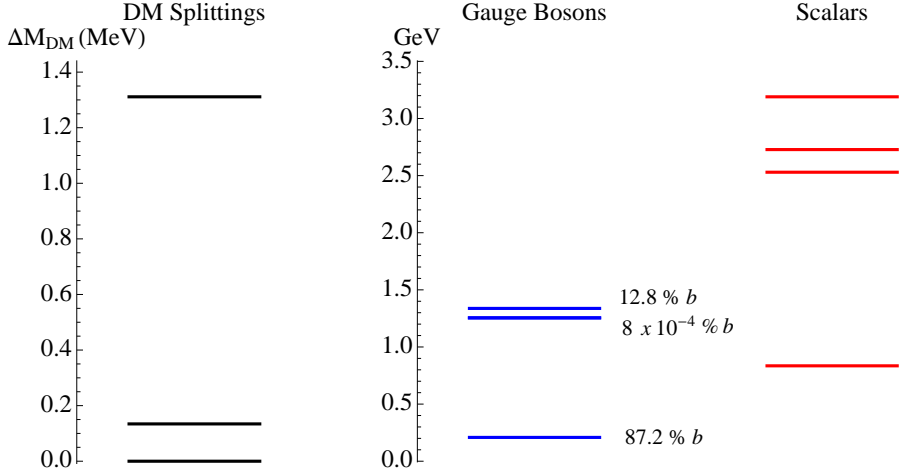}
\caption{\label{fig:2Dspec} The spectrum of {\bf Non-SUSY 1}, our
two doublet non-SUSY benchmark.  The left side shows the
radiative mass splittings of the components of the dark matter
triplet, measured from the ground state.  The splittings allow for the XDM and iDM explanations of INTEGRAL and DAMA, respectively.  The right side displays the
spectrum of the GeV-scale dark sector.  The $b$ fractions of the
gauge bosons are indicated and determine how strongly each gauge
boson couples to Standard Model electromagnetic current.  Because
of custodial $SU(2)$, two of the gauge bosons are degenerate and
do not mix with the $b$ at tree-level, and these are the gauge
bosons that couple between different dark matter states.  They do
mix with the $b$ at one-loop, inducing a suppressed iDM coupling,
and we include the dimension 6 operators $c^T_1 | h_1 D h_1  |^2$  and $c^T_2 | h_2 D h_2  |^2$ in order to parametrize custodial breaking
corrections.  The parameters of this benchmark are listed in the
text. }
\end {figure}

We begin with the two doublet model of Section \ref{subsect:2D},
where $h_1$ and $h_2$ have dark quantum numbers ${\bf 2}_{-1/2}$,
${\bf 2}_{1/2}$.  We have chosen a benchmark which breaks charge
and radiatively generates the XDM and iDM splittings, and we have calculated its mass spectrum
(Fig. \ref{fig:2Dspec}).  As discussed in section
\ref{sec:symmetrybreaking}, the custodial $SU(2)$ symmetry of the
Higgs sector determines the tree-level gauge boson spectrum.  The
gauge bosons that couple between the different dark matter mass
eigenstates, $w_\pm$, do not mix with the $b$ and are degenerate
in mass.  Custodial symmetry is broken at one-loop and in general
due to higher-dimensional operators.  The DAMA inelastic
scattering is therefore suppressed relative to models where
custodial symmetry is broken at tree-level.  For this benchmark,
we induce the iDM coupling by including the dimension 6
custodial-breaking operators $c^T_1 | h_1 D h_1  |^2$  and $c^T_2 | h_2 D h_2  |^2$, with coefficients $c^T_1$ and $c^T_2$ expressing the loop-suppression.

For the benchmark, we choose the gauge couplings: $g = 0.46$ and
$g_y = 0.19$.  The dark matter hypercharge is chosen to be $y_{\rm
dm} = 1/2$.  In the limit of small charge breaking, this choice
leads to one small and one large dark matter splitting, as
discussed in Section \ref{subsec:radiativesplitting}.  In terms of
the potential of Eq.~\ref{eqn:2HDPot}, the parameters are: $v_1 =
1.5 \unit {GeV}$, $v_2 = (1.5 + 3.2 \, i) \unit{GeV}$, $\lambda_1
= 0.5$, $\lambda_2 = 0.3$, $\lambda_3 = -0.031$, $\lambda_4 =
0.5$, and $\cos \alpha = 0.8$.  The coefficients of the
custodial-breaking dimension 6 operators are chosen to be
$c^T_1=2.8 \times 10^{-4} \, \GeV^{-2}$ and $c^T_2 =-5.7 \times
10^{-4} \, \GeV^{-2}$.

\subsubsection{Non-SUSY 2: Two Doublets and One Complex Triplet}
\label{subsubsec:nonSUSY2}

\begin {figure}[!h]
\hspace {-1.25cm} \includegraphics[width=1.2\textwidth]{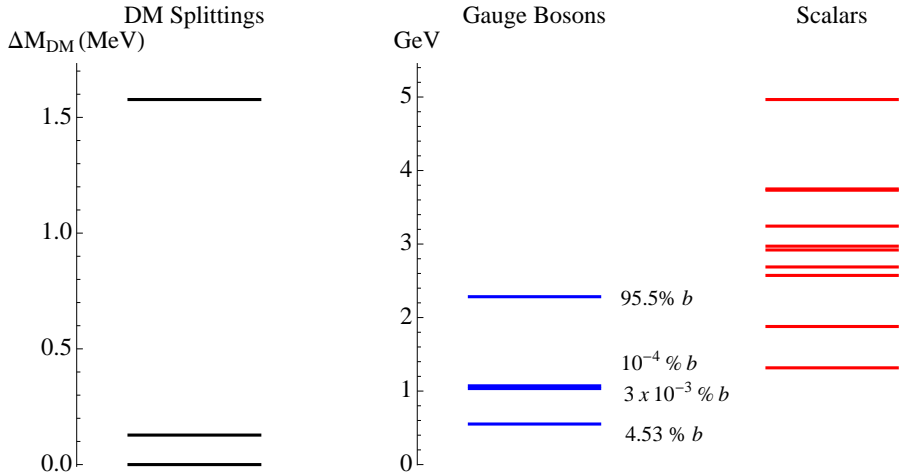}
\caption{\label{fig:2D1cTSpec}
The spectrum of {\bf Non-SUSY 2}, our two doublet and one complex triplet non-SUSY benchmark.  The left side displays the radiative mass splittings of the dark matter triplet, measured from the ground state.  The splittings can account for the XDM and iDM explanations of INTEGRAL and DAMA, respectively.  The right side shows the dark sector spectrum, and the $b$ fractions of the gauge bosons are indicated.  The triplet vev breaks custodial $SU(2)$, and all 4 gauge boson mass eigenstates mix with the $b$ at tree-level, although for this example most of the $b$ is contained in two of the mass eigenstates.  The parameters of this benchmark are listed in the text.
}
\end {figure}

We now add a complex triplet Higgs $\Phi$ to the two doublet model, with dark quantum numbers ${\bf 3}_0$.  The triplet vev breaks custodial symmetry, causing all gauge bosons to mix with the $b$ and inducing the iDM coupling at tree-level.  We take the triplet to be complex. While not the minimal possible choice, it has a more straightforward SUSY extension.  We again choose a benchmark that breaks charge in the doublet sector and radiatively generates the XDM and iDM dark matter splittings.  We have calculated its spectrum (Fig. \ref{fig:2D1cTSpec}).

For this benchmark, we choose the gauge couplings $g = 0.23$ and $g_y = 0.75$, and the dark matter hypercharge is chosen to be $y_{\rm dm} = 0.3$.  The potential is similar to Eq.~\ref{eqn:2D1TP} except we take $\Phi$ to be complex:

\begin{eqnarray} \label{eqn:2D1cTP}
&& \begin{tabular}{c||c|c|c}  field & $h_1$ & $h_2$ & $\Phi$
\tabularnewline \hline \hline charge & ${\bf 2}_{-1/2}$ & ${\bf
2}_{1/2}$ & ${\bf 3}_{0}$
\end{tabular} \\ \nonumber\\
V(h_1,h_2,\Phi) &=& V(h_1,h_2) +
\frac{\lambda_\Phi}{2}\left(\mathrm{Tr}\left[\Phi^\dagger \Phi
\right] - |v_\Phi|^2 \right)^2 \\\nonumber &+&  \left( c_1
h_1^\dagger \Phi h_1 + c_2 h_2^\dagger \Phi h_2 +c_3 h_1^T
\epsilon \Phi h_2 + \mathrm{h.c.} \right)
\end{eqnarray}
where the first term is the two doublet potential of Eq.~\ref{eqn:2HDPot}.  For the two doublet sector we choose the parameters: $v_1 = 1.8 \unit {GeV}$, $v_2 = (1.8 + 1.4 \, i) \unit{GeV}$, $\lambda_1 = 0.71 $, $\lambda_2 = 0.47$, $\lambda_3 = 0.33$, $\lambda_4 = 0.099$, and $\cos \alpha = 0.052$.  The parameters involving the triplet are chosen to be:  $v_\Phi = (1.1 + 0.61\,i) \unit{GeV}$, $\lambda_\Phi = 0.51$, $c_1 = (0.054 + 0.47 \, i) \unit{GeV}$, $c_2 = (0.74+0.69 \, i) \unit{GeV}$, and $c_3 = (0.61 + 0.81 \, i )\unit{GeV}$.

\subsection{SUSY Benchmark Models}
\label{subsec:SUSYbench}

Now we consider two SUSY benchmarks,  where the GeV scale is
generated radiatively from interactions with the Standard Model,
as discussed in Section \ref{sec:SUSY}.  Since our models all
employ a kinetic mixing, they all receive
 SUSY breaking contributions from kinetic mixing mediation, as
discussed in Section \ref{subsect:KMBreak}.  In our first example,
{\bf SUSY 1}, this the only source of SUSY breaking, however as
noted in appendix \ref{app:CBM}, it is difficult within this
framework to break charge with only one hypercharge neutral
triplet. We circumvent this in this example by adding a second
complex triplet and taking the triplets to have dark hypercharge.
For our second example, {\bf SUSY 2}, we add an additional
gauge mediation source for GeV-scale SUSY breaking by taking dark
matter to be charged as a $\mathbf{5} + \mathbf{\bar 5}$ of the SM\@. Dark matter then
acts as a messenger of gauge mediation, as discussed in section
\ref{subsect:55barMess}.  For this setup, we can break charge with
two doublets and one hypercharge neutral triplet.

\subsubsection{SUSY 1: Kinetic Mixing Mediation with Two Doublets and Two Triplets}
\label{subsubsec:SUSYbench1} For this benchmark, we have two
doublets, $h_1$ and $h_2$, and two complex triplets, $\Phi_1$ and
$\Phi_2$, with dark quantum numbers ${\bf 2}_{-1/2}$, ${\bf
2}_{1/2}$, ${\bf 3}_1$, and ${\bf 3}_{-1}$.  We have chosen
triplet hypercharge assignments that allow Yukawa couplings
between doublets and triplets, otherwise there may be
pseudo-Goldstone bosons in the spectrum.  The GeV scale is
generated in the dark sector from kinetic mixing, as described in
Section \ref{subsect:KMBreak}. The most general renormalizable
superpotential for two doublets and two triplets with these charge
assignments is:
\be \label{eqn:W2D2T} && \begin{tabular}{c||c|c|c|c} field & $h_1$
& $h_2$ & $\Phi_1$ & $\Phi_2$ \tabularnewline \hline \hline charge
& ${\bf 2}_{-1/2}$ & ${\bf 2}_{1/2}$ & ${\bf 3}_{1}$ & ${\bf
3}_{-1}$
\end{tabular} \\ \nonumber\\
W &=& \mu_h h_1^T \epsilon h_2 + \mu_\Phi
\mathrm{Tr} \left[ \Phi_1 \Phi_2 \right] + \lambda_1 h_1^T
\epsilon \Phi_1 h_1 +\lambda_2 h_2^T \epsilon \Phi_2 h_2 \ee
We include GeV scale $\mu$ and $B\mu$ terms for the doublets and
triplets because they help break the dark gauge symmetry and lift
runaway directions.  We do not include the effects of running from
the TeV scale to the GeV scale, which we take to be subdominant.
Kinetic mixing mediation already gives negative scalar soft mass
squareds at tree-level, leading to the breaking of dark gauge
symmetry.

\begin {figure}[!h]
\hspace {-1.25cm}
\includegraphics[width=1.2\textwidth]{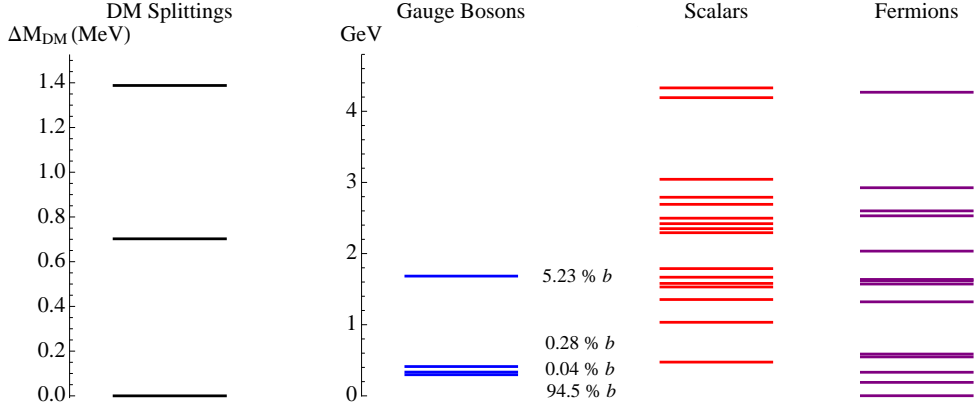}
\caption{\label{fig:2D2Tsusyspec} The spectrum of {\bf SUSY 1},
our two doublet and two complex triplet SUSY benchmark with
kinetic mixing mediation.  The left side displays the radiative
mass splittings of the dark matter triplet, measured from the
ground state.  The larger splitting allows for the XDM explanation of INTEGRAL, but the
smaller splitting is too large to explain DAMA with iDM\@.  The right side
shows the dark sector spectrum, and the $b$ fractions of the gauge
bosons are indicated.  The triplet vevs break custodial symmetry,
and all four gauge boson mass eigenstates are part $b$ at
tree-level.  The gauginos and Higgsinos are strongly mixed after
dark symmetry breaking.  The lightest fermion, with mass $\sim 2
\unit{MeV}$, is mostly gaugino and light because gauginos get no
soft masses from kinetic mixing mediation.  The second lightest
fermion has a 3-body decay to the lightest fermion and a SM lepton
pair, which can account for astrophysical lepton production and
lead to a visibly displaced vertex at colliders. }
\end {figure}

We include a kinetic mixing coefficient of $\epsilon = 2 \times
10^{-4}$, in terms of Eq.~\ref{eq:kmix}, which automatically
generates the GeV scale in the dark sector.  Our benchmark
radiatively generates the XDM splitting, but unfortunately the
smaller dark matter splitting is too large to account for iDM\@.  We
have calculated the mass spectrum (Fig. \ref{fig:2D2Tsusyspec}).
The triplet vevs break custodial $SU(2)$ at tree-level and all
gauge bosons mix with the $b$.  The gauginos and Higgsinos are
strongly mixed after dark symmetry breaking, but for this example
the lightest fermion is a mostly gaugino-like Goldstino with a
mass of only $\sim 2 \unit{MeV}$.  Such a field is present because
SUSY is broken within the dark sector itself. The
second lightest fermion, with mass $\sim 190 \unit{MeV}$, is
lighter than the lightest gauge boson. Thus, the dark gauge bosons
will cascade into these light fermions, rather than SM lepton
pairs.  The second lightest fermion decays to the lightest fermion
and a SM lepton pair through a 3-body decay, which can account for
the astrophysical lepton production and lead to visibly displaced
vertices at colliders.  Another possibility, not realized in this
example, is to have an approximately supersymmetric dark sector,
with a kinetic mixing mediation induced super-Higgs mechanism at a
GeV\@. Gauginos then reside in massive vector supermultiplets and
get GeV scale masses.

For this benchmark, we have chosen the gauge couplings $g = 0.22$
and $g_y = 1.2$, and dark matter hypercharge $y_{\rm dm} = 1/5$.
The superpotential Yukawa couplings are chosen to be $\lambda_1 =
1.7 + 0.022 \,i$ and $\lambda_2 = 0.5 + 1.8 \, i$.  For the
doublets we choose $\mu_h = (0.11 + 0.63 \, i) \unit {GeV}$ and
$(B\mu)_h = (0.74 + 0.69 \, i) \unit {GeV}^2$.  For the triplets
we choose $\mu_\Phi = (0.51 + 0.83 \, i) \unit{GeV}$ and
$(B\mu)_\Phi = (0.57 + 0.59 \, i) \unit{GeV}^2$.

\subsubsection{SUSY 2: Gauge Mediation with Two Doublets and One Triplet}
\label{subsubsec:SUSYbench2}

For this benchmark, we supersymmetrize the Higgs content of our
{\bf Non-SUSY 2} benchmark, including SUSY breaking contributions
from both kinetic mixing mediation and gauge mediation with dark
matter messengers. The most general renormalizable superpotential
for two doublets and one triplet is the following:
\be \label{eqn:W2D1T}&& \begin{tabular}{c||c|c|c} field & $h_1$ &
$h_2$ & $\Phi$ \tabularnewline \hline \hline charge & ${\bf
2}_{-1/2}$ & ${\bf 2}_{1/2}$ & ${\bf 3}_{0}$
\end{tabular} \\ \nonumber\\ W &=& \mu_h h_1^T \epsilon h_2 + \mu_\Phi
\mathrm{Tr} \left[ \Phi^2 \right] + \lambda h_1^T \epsilon \Phi
h_2 \ee
As in the {\bf SUSY 1} benchmark, we include GeV scale $\mu$ terms
for the doublets and triplet and do not include the effects of
running from the TeV scale to the GeV scale.  There are already
negative scalar soft mass squareds at tree-level  because of the
nonzero dark matter supertrace \cite{Poppitz:1996xw}, leading to
the breaking of dark gauge symmetry.  For this example, it is not
necessary to include GeV scale $B\mu$ terms for the doublets or
triplet.

\begin {figure}[!h]
\hspace {-1.25 cm} \includegraphics[width=1.2\textwidth]{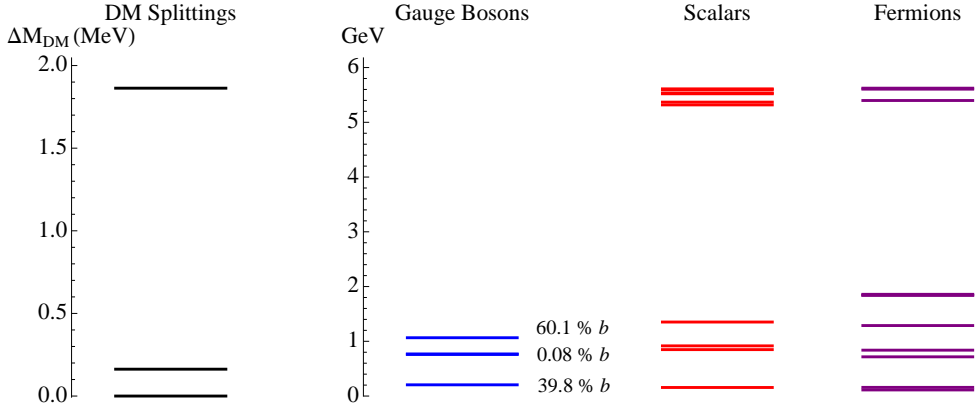}
\caption{\label{fig:2D1Tsusyspec}
The spectrum of {\bf SUSY 2}, our two doublet and one complex triplet SUSY benchmark with both kinetic mixing mediation and dark matter messengers charged under the Standard Model.  The left side displays the radiative mass splittings of the dark matter triplet, measured from the ground state.  The splittings can allow for the XDM and iDM explanations of INTEGRAL and DAMA, respectively.  The right side shows the dark sector spectrum, and the $b$ fractions of the gauge bosons are indicated.  The triplet vev breaks custodial $SU(2)$ and three of the gauge boson mass eigenstates are part $b$ at tree-level.  The dark spectrum now includes GeV scale fermions, and the gauginos and Higgsinos are strongly mixed after dark symmetry breaking.  The three heaviest fermions, with masses near $\sim 5.5 \unit{GeV}$,  are almost pure Higgsino mixtures.  The spectrum is slightly split by a small separation between the dark $\mu$ and soft mass scales.   The parameters of this benchmark are listed in the text.
}
\end {figure}

We have chosen a benchmark which generates a GeV scale dark sector
with charge breaking and custodial breaking, and which leads to
radiative XDM and iDM splittings.  We have calculated the mass
spectrum (Fig. \ref{fig:2D1Tsusyspec}).  The gauginos and
Higgsinos are strongly mixed after dark symmetry breaking, but the
three heaviest fermions with masses near $\sim 5.5 \unit{GeV}$ are
almost pure Higgsino mixtures.  The spectrum is slightly split by
a small separation between the dark $\mu$ and soft mass scales.
The gauge couplings are chosen to be $g = 0.3$ and $g_y = 0.37$
and the dark matter hypercharge is  $y_{\rm dm} = 1 / 2$.  The
kinetic mixing is $\epsilon = 7 \times 10^{-5}$ in terms of
equation \ref{eq:kmix}.  The messenger scale of SUSY breaking to
the Standard Model, which enters the log divergence of
Eq.~\ref{eqn:55barMs}, is chosen to be $\Lambda_{UV} = 30
\unit{TeV}$, corresponding to low-scale gauge mediation.   The
standard model doublet dark matter mass components, in terms of
Eq.~\ref{eq:dmmassmat}, are given by $m_f^{({\bf 2})} = 800
\unit{GeV}$, $m_s^{({\bf 2})} = 50 \unit{GeV}$, and $B\mu^{({\bf
2})} = (40 \unit{GeV})^2$.  As discussed in section
\ref{subsect:55barMess}, the small soft mass is needed to generate
the GeV scale in the dark sector, and we choose a small $B\mu$,
keeping dark matter fermionic.  For the standard model triplet
components of the dark matter $\mathbf{5} + \mathbf{\bar 5}$, we
choose the parameters $m_f^{({\bf 3})} = 840 \unit{GeV}$,
$m_s^{({\bf 3})} = 50 \unit{GeV}$, and $B\mu^{({\bf 3})} = (300
\unit{GeV})^2$.  The larger $B\mu$ for the triplet leads to GeV
scale gaugino soft masses in the dark sector (see
Eq.~\ref{eqn:55barMlambda} and the surrounding discussion).  The
superpotential parameters are $\mu_h = (0.27+0.28\,i) \unit{GeV}$,
$\mu_\Phi = (2.52 + 3.48 \, i) \unit{GeV}$, and $\lambda = 0.29 +
1.51 \, i$.

\section{Signals of a Non-Abelian Dark Sector at the Tevatron and the LHC}
\label{sec:collider}

 In this section we discuss the collider phenomenology associated
with the models presented in the previous sections. In the first
part of this section we analyze the generic predictions associated
with a non-abelian dark sector that is linked to the SM only via
kinetic mixing. In the second part we present the signals expected
in supersymmetric versions of such models. Throughout, we limit
the discussion to the Tevatron and LHC\@.
It is important to realize that in the case of a GeV scale dark sector such high-energy accelerators are needed more for their luminosity than their energy reach, which is considerably higher than the dark sector scale. In supersymmetric implementations, colored MSSM superpartners can be copiously produced at hadron colliders. Their subsequent decay into dark states can produce spectacular signals involving multiple lepton jets.  We leave it for future work to investigate the
phenomenology of these models at low-energy experiments, but see
Ref.~\cite{Pospelov:2008zw} for low-energy signatures of similar
models.

A new sector of light particles with very weak couplings to the Standard Model
have been discussed in detail in the context of the ``Hidden
Valley" models \cite{Strassler:2006im}. Their collider phenomenology was investigated in \cite{Han:2007ae,Strassler:2008fv}. In particular, the modifications such models can introduce to the
decay chains of the MSSM was clarified in Ref.~\cite{Strassler:2006qa}. Here,
we focus on the particular scenario which uses the kinetic mixing as the essential link
between the SM and the dark sector. In addition, motivated by astrophysical observations, we allow the dark sector to decay back to light leptons ($e^\pm$ and $\mu^\pm$) only. For the purpose of this paper, we do not concern ourselves with a possibly small branching fraction into pions.

The pair production of the dark matter states at colliders is certainly possible if they happen to carry SM weak charge. However, their detection proves extremely difficult since they are not accompanied by any hard object. Even if the excited states of its SM multiplet are produced, their decays are too soft to trigger on since they are separated by only $\sim\GeV$ (notice that this splitting is generated by the SM gauge interactions and are of order $\alpha M_{Z}$ \cite{Thomas:1998wy}).

\subsection{Production and decay of dark gauge bosons and Higgses}

As discussed in detail in appendix~\ref{app:KinMixing}, the kinetic mixing induces two
important, $\epsilon$ suppressed, couplings: The SM electromagnetic current is now also charged under the dark gauge bosons;  the SM $Z^0$ boson is now coupled to the dark hypercharge current. Before discussing each of these couplings and their impact on collider signals, let us
briefly discuss the decays of the dark gauge boson and the dark Higgses.

\subsubsection{Dark gauge boson and Higgs decay chains}
  \label{subsubsec:lepjetstruc}

\begin{figure}[h!]
\begin{tabular}{cccc}
\parbox{3.0cm}{\includegraphics[scale=0.4]{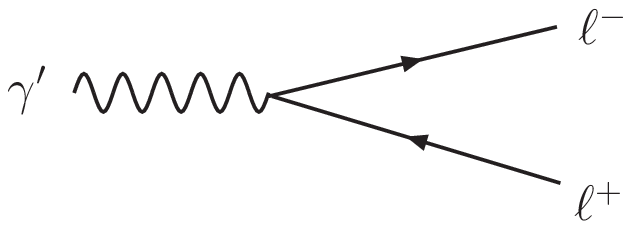}} &
\parbox{3.0cm}{\includegraphics[scale=0.4]{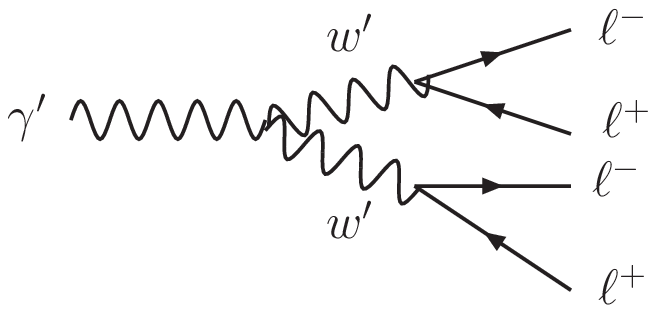}} &
\parbox{3.5cm}{\includegraphics[scale=0.4]{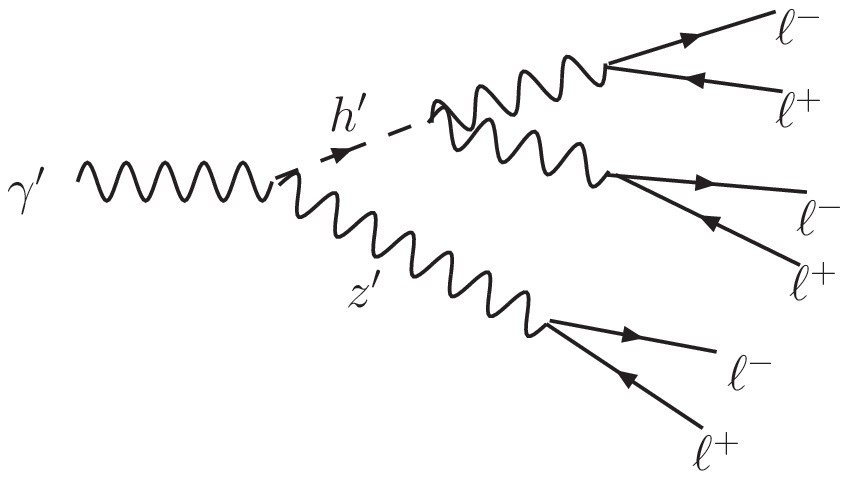}} &
\parbox{3.5cm}{\includegraphics[scale=0.4]{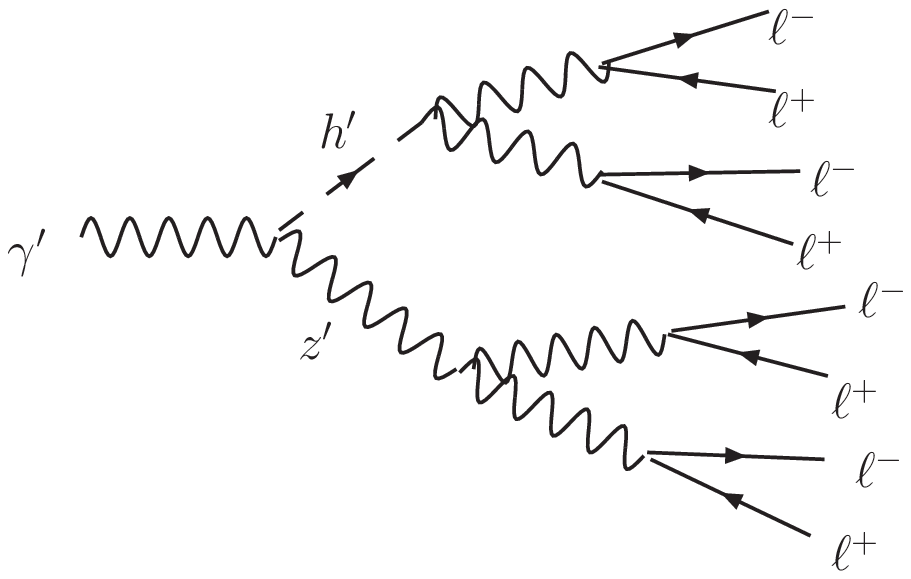}}
\end{tabular}
\caption{\label{fig:non-susy-decay} Typical decay chains starting with
a dark gauge boson, labelled $\gamma'$ in this plot. The dark decay chain can have several stages and involve additional dark sector states, such as other dark gauge bosons (labeled $w', z'$ in this figure), and dark Higgses (labelled $h'$). }
\end{figure}

The non-abelian nature of the dark sector implies the presence of
complicated decay chains. Some of the typical decays chains are
shown in Fig.~\ref{fig:non-susy-decay}.
In the dark sector, gauge boson mass eigenstates are generically
mixtures of all four $SU(2)_{\rm dark}\times U(1)_y$ gauge
eigenstates. In Fig.~\ref{fig:non-susy-decay}  and the rest of
this section, we have used $\gamma'$ (and also $w'$ and $z'$ in
this figure) to denote any one of these mass eigenstates. For an
abelian dark sector with kinetic mixing with the SM,
$\gamma'$ decay leads to a di-lepton final state, shown in the
first panel from the left of Fig.~\ref{fig:non-susy-decay}.  On
the other hand, a non-abelian dark sector, like one of the
examples considered in this paper, leads to complicated decay
chains, such as the ones shown in the rest of
Fig.~\ref{fig:non-susy-decay}. The dark Higgs sector, necessary to break the non-abelian group, may also participate in such cascades as shown in the right two panels of
Fig.~\ref{fig:non-susy-decay}. Such cascades inevitably
produce multiple, easily $> 2$ and possibly 8, final state
leptons, which provides a unique signature of the non-abelian
nature of the dark sector\footnote{Sometimes phase space
constrains the flavor of the lepton. For example, a GeV dark gauge
boson cannot decay into more than 4 muons}. We expect the decay between dark states to be
generically prompt. Therefore, the decay length is dominated by the very
last decays back into SM leptons. A rough estimate for a generic
decay is then,
\begin{equation}
\label{eqn:decaylength}
c\tau^{\gamma'\rightarrow {\rm n} \ell}_{\rm
2-body} \sim \frac{1}{\alpha \epsilon^2 m_{\gamma'} }= 2.7 \times
10^{-6} \unit{cm} \; \left(\frac{\mbox{GeV}}{m_{\gamma'}}\right)
\left(\frac{10^{-3}}{\epsilon}\right)^2.
\end{equation}
With moderate boost $\gamma \sim \mathcal{O}(10)$, this may lead to a displaced vertex
if $\epsilon \lesssim 10^{-4}$.

To be observable at hadron colliders, the dark boson which initiates such a cascade must carry $p_{\rm T} \sim {\mathcal{O}} (10{\rm s})$ GeV\@. Therefore, regardless of the precise nature of the cascade which ensues, its decay products have small opening
angles $\delta \theta \sim m_{\gamma'} / p_{\rm T} < 0.1$. Those decay products will eventually decay into several collimated SM leptons. A collection of more than 2 hard and collimated leptons is dubbed a
``lepton jet'' \cite{ArkaniHamed:2008qp}.

\subsubsection{Displaced vertices and missing energy}
\label{subsubsec:MET}

While Eq. (\ref{eqn:decaylength}) is the generic estimate for the resulting decay length of dark cascades, there are several exceptions which may result in more noticeably displaced vertices or missing energy in lepton jets.

If it is kinematically forbidden for a dark gauge boson to have
2-body on-shell decays within the dark sector, then the dark gauge
boson may decay directly into two leptons. However, a noticeable
exception occurs when the 3-body decay $\gamma' \rightarrow a'^*
b'_1 \rightarrow b'_1 b'_2 b'_3 $ is kinematically allowed, where
$a'$, $b'$ can be either dark gauge boson or dark Higgs states. In
this case, there is an additional suppression of $(\delta m /
m_{a'})^5 \times \mbox{(3-body \ phase space)}$  on the decay
width, where $\delta m \sim m_{\gamma'}-\sum_i m_{b'_i}$, and we
have used $m_{\gamma'} \sim m_{a'}$ in this estimate. This decay
channel can be competitive and even dominate over the direct decay
into 2 leptons. In particular, when the decay into SM leptons is strongly suppressed (dark pseudoscalar decay) or all together forbidden (dark fermion decay), the 3-body process may dominate and lead to a displaced vertex. The impact parameters of
multiple leptons associated with this displaced vertex will not be
correlated with each other since they come from the decays of
different resonances $b'_{1,2,3}$.

\begin{figure}[h!]
\begin{center}
\includegraphics[scale=0.45]{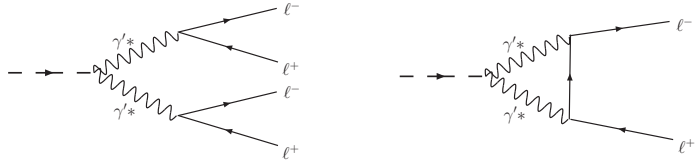}
\caption{\label{fig:pseudo-decay} Two possible decay channels if the
  lightest dark sector state is a scalar from the dark Higgs sector. }
\end{center}
\end{figure}

If the lightest dark sector state is a dark Higgs, $h'_0$, it
cannot directly decay into SM leptons (unless it mixes the SM
Higgs, see Ref.~\cite{Finkbeiner:2007kk,ArkaniHamed:2008qn}). In
this case, the dark Higgs will either decay into 4 leptons through
two off-shell gauge bosons, shown in the left panel of
Fig.~\ref{fig:pseudo-decay}, or into 2 leptons through a one-loop
decay. Either way, such a decay leads to a very long life-time,
$c\tau \sim {\mathcal O}(\mbox{km})$ for $m_{h_0'} \lesssim\GeV$.
In this case, dark cascades which involve this lightest scalar
contain missing energy as it escapes the detector. These cascades
can still produce observable lepton jets because, in addition to
missing energy, one still gets leptons from the intermediate steps
of the decay, such as $h'_i \rightarrow a' h'_0$ followed by $a'
\rightarrow$ lepton pairs. In this case, the lepton jet contains
missing energy that is collimated with the leptons of the same
cascade.

An additional source of missing energy comes in a supersymmetric dark
sector with R-parity. The lightest dark supersymmetric particle
(LDSP) may be stable if the gravitino is heavier. Otherwise, it
may eventually decay into the gravitino. Either way, it carries
with it missing energy. Unless the MSSM sector decays directly
into the LDSP, in which case there may be no lepton jets, missing
energy due to the LDSP will be collimated with the visible
lepton jets, very similarly to the non-SUSY case. Such correlations
provide an additional handle on the reconstruction of these events
since we know the direction of the missing particles and
can treat them as having vanishing masses. We provide an example
of such a reconstruction in the case of rare $Z^0$ decays below.

\subsubsection{Direct Production}
\label{subsubsec:direct-prod}

\begin{figure}[h!]
\begin{center}
\includegraphics[scale=0.55]{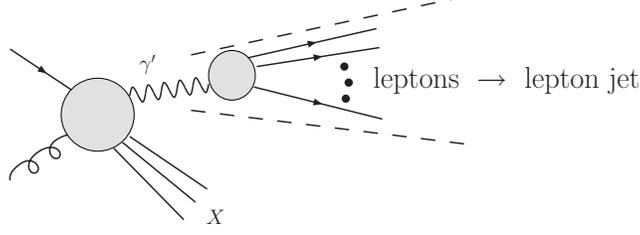}
\caption{\label{fig:gammaXproduction} Direct production of a dark gauge
boson in a process very similar to prompt photon production in
the Standard Model. }
\end{center}
\end{figure}

\begin{figure}[h!]
\begin{center}
\includegraphics[angle=270,scale=0.27]{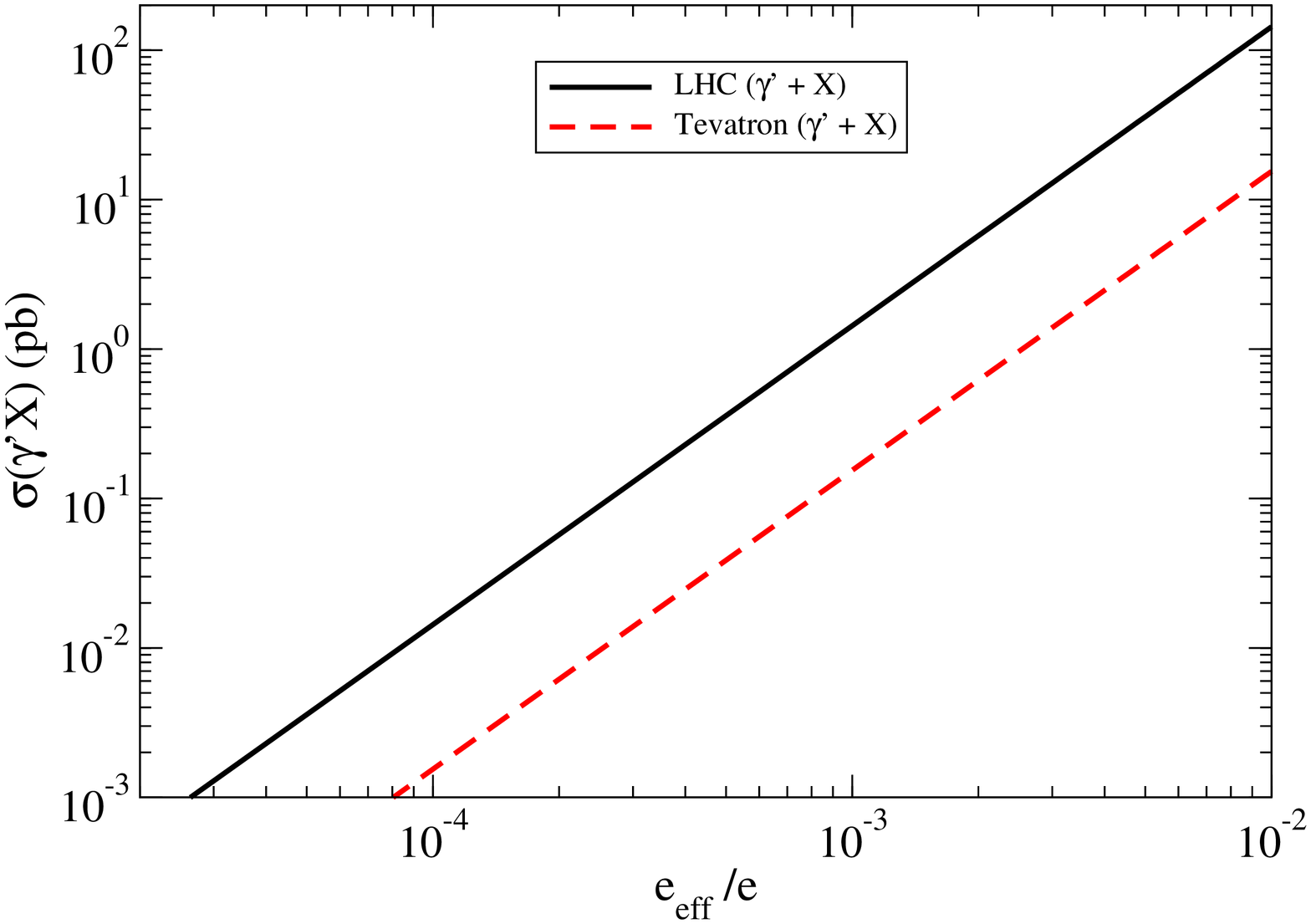}
\includegraphics[angle=270,scale=0.27]{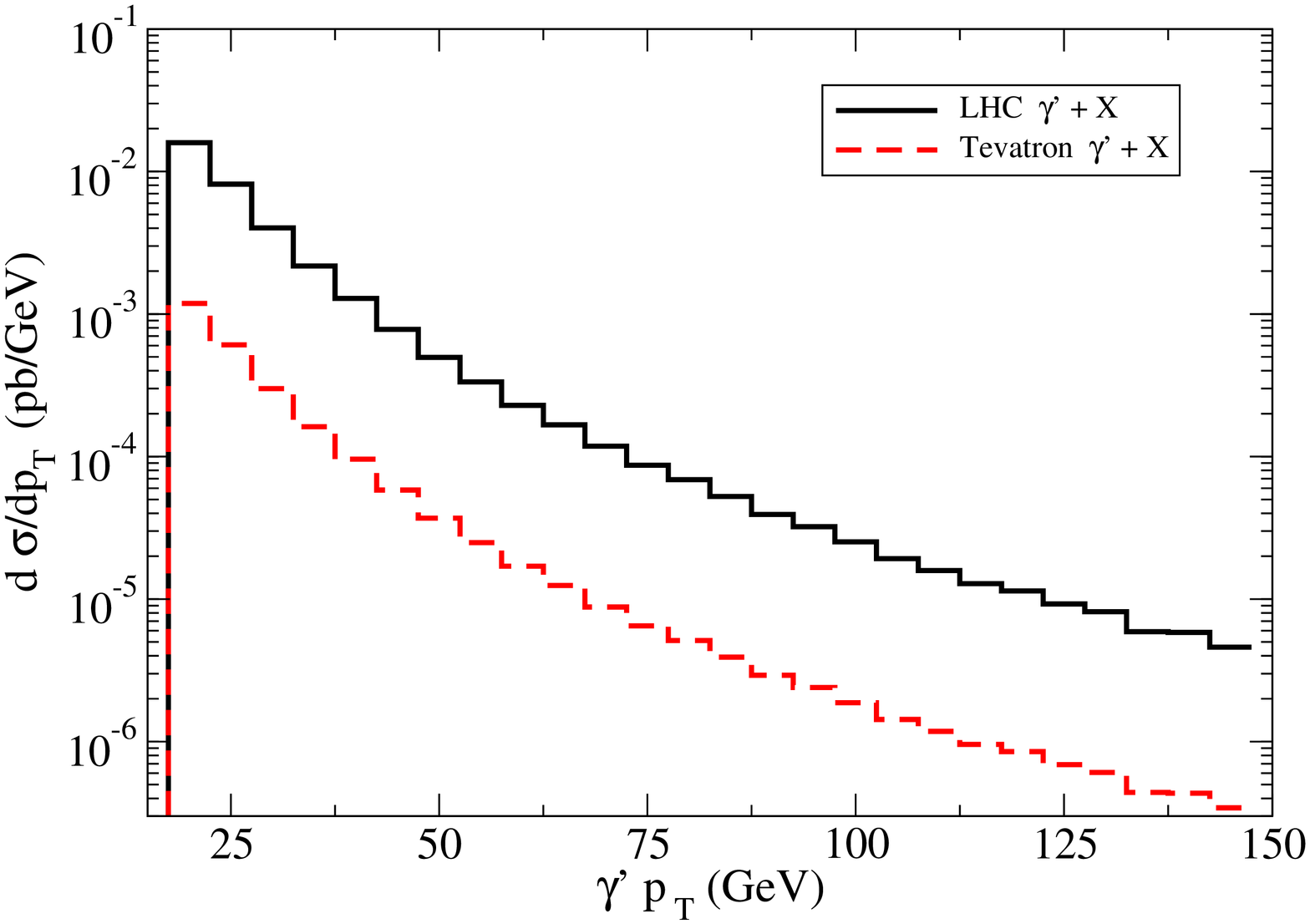}
\caption{\label{fig:gammaXrate} In the left pane, we show the rate
of direct production of the dark gauge boson as a function of
$e_{\rm eff}/e$, where $e_{\rm eff}$ is the effective coupling of
dark gauge boson to the Standard Model fields. }
\end{center}
\end{figure}
\vspace{5mm}

The kinetic mixing between the dark force carrier and the SM photon induces a small dark
charge for electromagnetically charged SM fields. Consequently, the dark gauge boson can be
directly produced in colliders via a process analogous to prompt
photon production in the SM, shown in Fig.~\ref{fig:gammaXproduction}.

In the left panel of Fig.~\ref{fig:gammaXrate}, we present the
production rate of dark gauge bosons as a function of $e_{\rm
eff}/e$, where $e_{\rm eff} =\epsilon e \cos \theta_{W} f_b$ is
their effective gauge coupling to SM fields\footnote{The
simulation was actually of prompt photon production with PYTHIA
\cite{Sjostrand:2006za} and the resulting cross-section was
multiplied by a factor of  $e_{\rm eff}^2/e^2$.} and $f_b$ is the
fraction of the dark hypercharge gauge boson $b_{\mu}$ in a given
dark gauge boson mass eigenstate. In the right panel of
Fig.~\ref{fig:gammaXrate} we plot the inclusive differential
cross-section of dark photon ($\gamma^\prime$) production at the
LHC and the Tevatron with $e_{\rm eff}=10^{-3} e$.

\begin{figure}[h!]
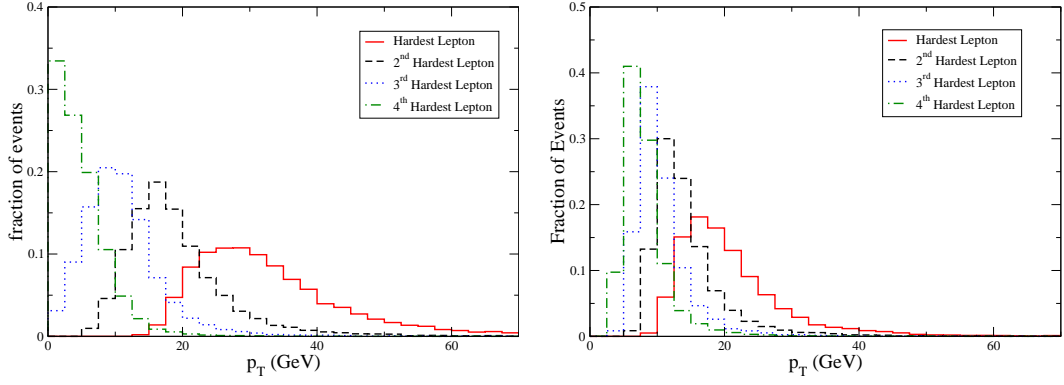

\begin{center}
\includegraphics[scale=0.29]{LepPT4l.eps} \  \
\includegraphics[scale=0.29]{LepPT8L.eps}
\end{center}
\caption{\label{fig:LepPT} $p_T$ distributions for cascades resulting in 4 (left) and 8 (right) leptons,  for events with $p_T>50\GeV$ for $\gamma'$.}
\end{figure}

\begin{figure}[h!]
\begin{center}
\includegraphics[scale=0.45]{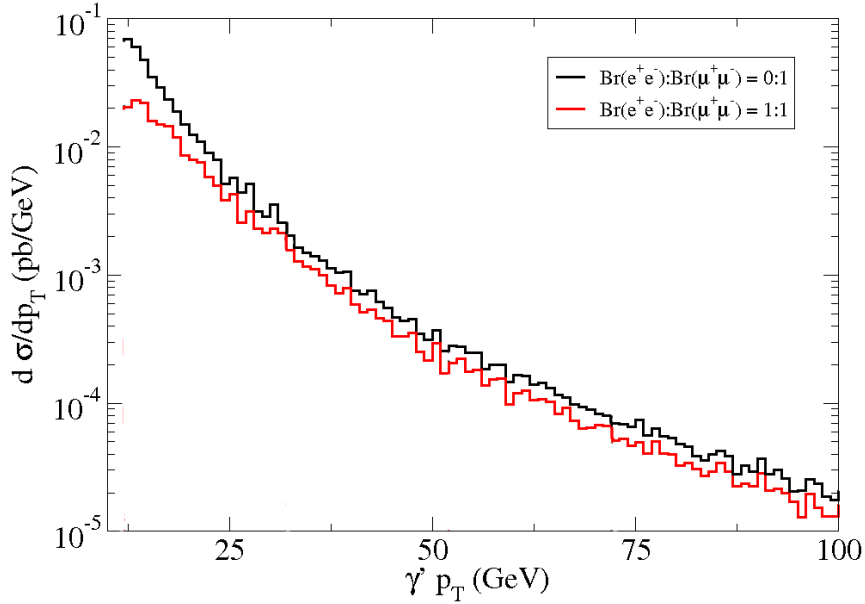}
\end{center}
\caption{\label{fig:gammaPTtrigX} The differential cross-section as a function of the $p_T$ of
  $\gamma'$ at the LHC ($\sqrt{s}=14\TeV$) after including muon
  triggers, demanding either a single muon with $p_T>7\GeV$ or two
  muons with $p_T>3\GeV$. Proper $\eta$ cuts were imposed and each
  event was required to contain at least 3 leptons.}
\end{figure}

After dark vector bosons are produced, they typically cascade down
to multiple leptons that form a lepton jet as discussed above.
These leptons carry a significant amount of $p_T$, as shown in
Fig.~\ref{fig:LepPT}. At CMS, the Level 1 Dimuon trigger (2 muons
with $p_T>3\GeV$ in $|\eta|<2.4$) or single muon trigger (1 muon
with $p_T>7\GeV$ and $|\eta|<2.4$) should be able to detect those
events that contain muons \cite{Ball:2007zza}. The electron
triggers are single e (isolated $E_T>26\GeV$), double e (isolated
$E_T>14.5$) and double relaxed e (not isolated $E_T>21.8\GeV$).
Since the resulting electrons are unlikely to be isolated
``electromagnetic'' objects, the double relaxed e is probably
necessary.  We will conservatively assume that muons alone are
triggered on. In Fig.~\ref{fig:gammaPTtrigX}, we show the
differential cross section of dark $\gamma'$, taking into account
the simple requirements on muon triggering.

\vspace{5mm}
\begin{figure}
\begin{center}
\includegraphics[scale=0.4]{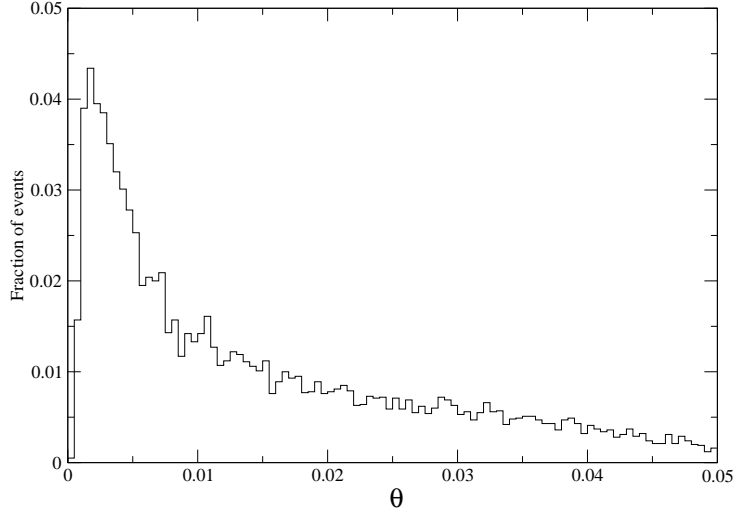}
\end{center}
\caption{\label{fig:maxTheta} The fraction of events with a maximum opening angle $\theta$ between leptons in a lepton jet, which contains 4 leptons.}
\end{figure}

\subsubsection{Distinguishing Leptons}

Let us discuss the issue of discriminating individual
leptons within a given lepton jet\footnote{We are grateful for
valuable discussions with Jim Olsen on this subject.}. In our
present discussion, we focus on muons.
 In Fig.~\ref{fig:maxTheta} we plot
the maximal opening angle between any two of the four leptons. At
such high momenta, the resulting decay products are highly
collimated with an initial opening angle of approximately $\theta
\sim m_{\gamma^\prime}/p_T < 0.1$, which can be as small as
$10^{-2}$. By the time these muons reach the first layer of the
muon system, they typically acquire a sufficient separation to be
distinguished. For example, as depicted in Fig.~\ref{fig:LepPT}, a
typical scenario will have two muons with average $p_T \sim 20$
GeV and $\Delta p_T \sim 5$ GeV\@. Without even including the
initial lepton jet opening angle, we estimate that the acquired
separation is about 10 cm (in the CMS detector), which is greater
than the cell size of $\sim 4.5$ cm. The separation between two
same sign muons is proportional to $\Delta p_T / (p_T^{\mu_1}
p_T^{\mu_2}) $. For a given lepton jet $p_T$, since both $\Delta
p_T$ and $p_T$ are inversely proportional to the number of
leptons, higher multiplicities actually result in larger
separations.   We also notice from Fig.~\ref{fig:LepPT} that
leptons in lepton jets typically have different $p_T$, such
that $\Delta p_T / p_T \sim 20 \%$ or so.  LHC detectors can
achieve better muon momentum resolution. For example, CMS can
achieve $\Delta p_T / p_T \sim 1 \% $ (about $10 \%$ with muon
system only) in the momentum regime of interest \cite{Bayatian:2006zz}. ATLAS can achieve a similar precision
\cite{ATLAS:1997fv}. Finally, the muon isolation separation defined by
CMS can be as small as $\Delta R = 0.01$. The angular resolution
is even smaller, about 2 mrad\cite{Bayatian:2006zz}\footnote{This
is the resolution quoted for a single hit. It is beyond our abilities to
evaluate how the resolution deteriorates with multiple collimated
muons.}. Thus, it is reasonable to assume that CMS will be capable of
resolving several, if not all of the muons. The primary background
arises from $K$ and $\pi$ decays and $J/\psi \rightarrow
\mu^+\mu^-$ (muons coming off soft jets can be vetoed with
isolation cuts), and possibly from other heavy flavor decays. The
high lepton multiplicity in those events and the lack of hadronic
activity around the lepton jet should be sufficient to fight these
backgrounds and obtain a clean sample. However, a more careful
collider analysis is certainly warranted, but is beyond the scope
of the present work.

\subsubsection{Rare $Z^0$ decay}
\label{subsubsec:rareZ}

\begin{figure}[h!]
\begin{center}
\includegraphics[scale=0.6]{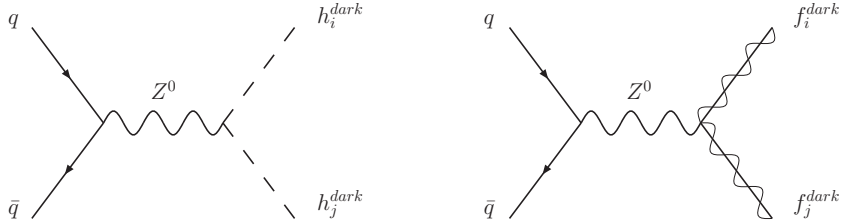}
\caption{\label{fig:z-prod-decay}$Z^0$ production and two possible decay channels into the dark sector. On the left we depict a decay into the dark Higgs sector. Fermionic channels, such as the one shown on the right, are dominantly associated with the Higgsino states possible in supersymmetric versions of the model.}
\end{center}
\end{figure}

As discussed in Appendix~\ref{app:KinMixing}, the kinetic mixing
also induces a coupling $\epsilon Z_{\mu} J^{\mu}_{b}$, where
$J^{\mu}_b$ is the dark hypercharge current.  Thus, we can produce
dark hypercharged states through rare decays of
the $Z^0$, shown schematically in Fig.~\ref{fig:z-prod-decay}. The $\epsilon^2$ suppression makes LEP searches irrelevant due to luminosity limits, but the Tevatron and LHC may probe such events. The decay branching ratio to any particular dark sector state $d_i$ can be written as
\begin{equation}
\mbox{BR} (Z^0\rightarrow d_ i d_i) = \frac{c_{d_i} }{\Gamma_Z^0} \frac{\epsilon^2 g_y^2  y_{d_i}^2\sin^2 \theta_W}{48 \pi} M_{Z^0} ,
\end{equation}
where $c_{d_i}$ depends on decay matrix element and is proportional to the number of degrees of freedom of $d_i$. The total branching ratio into the dark sector will scale linearly with the number of dark sector states, which could be easily $\mathcal{O}(10)$ in our scenario.

The SM photon does not couple to the dark sector states. However, there is a ``continuum'' contribution to the same amplitude through off-shell dark photon, $q \bar{q} \rightarrow \gamma'^{\star} \rightarrow d_i d_i$, which is proportional to $e_{\rm eff}^2 \propto \epsilon^2$. Depending on the spectrum and couplings in the dark sector, it could have important contributions to the signal off the Z-peak. In this section, we will focus on the contribution within the Z resonance.

\begin{figure}[h!]
\begin{tabular}{cc}
\includegraphics[angle=270,scale=0.3]{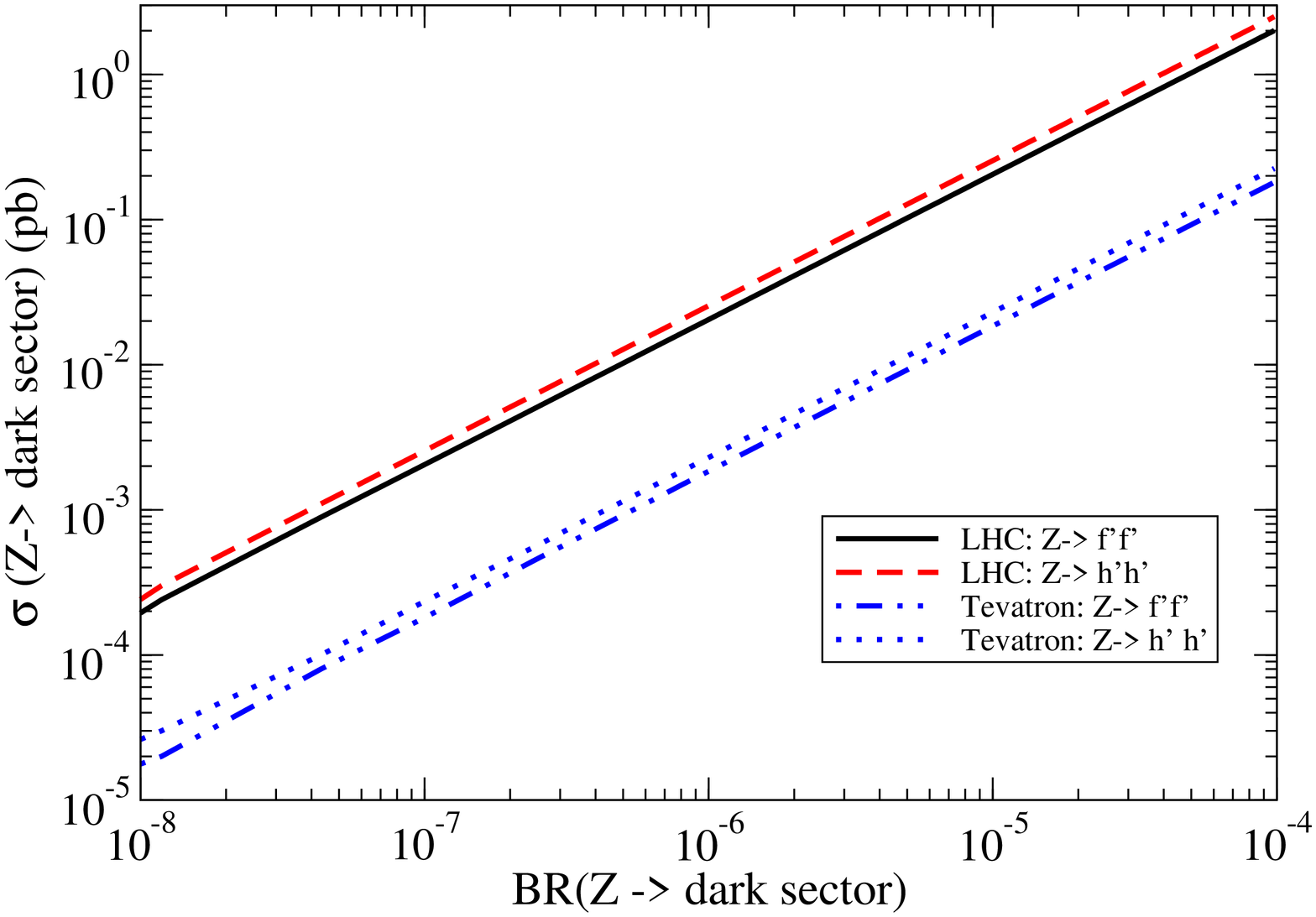}&
\includegraphics[angle=270,scale=0.3]{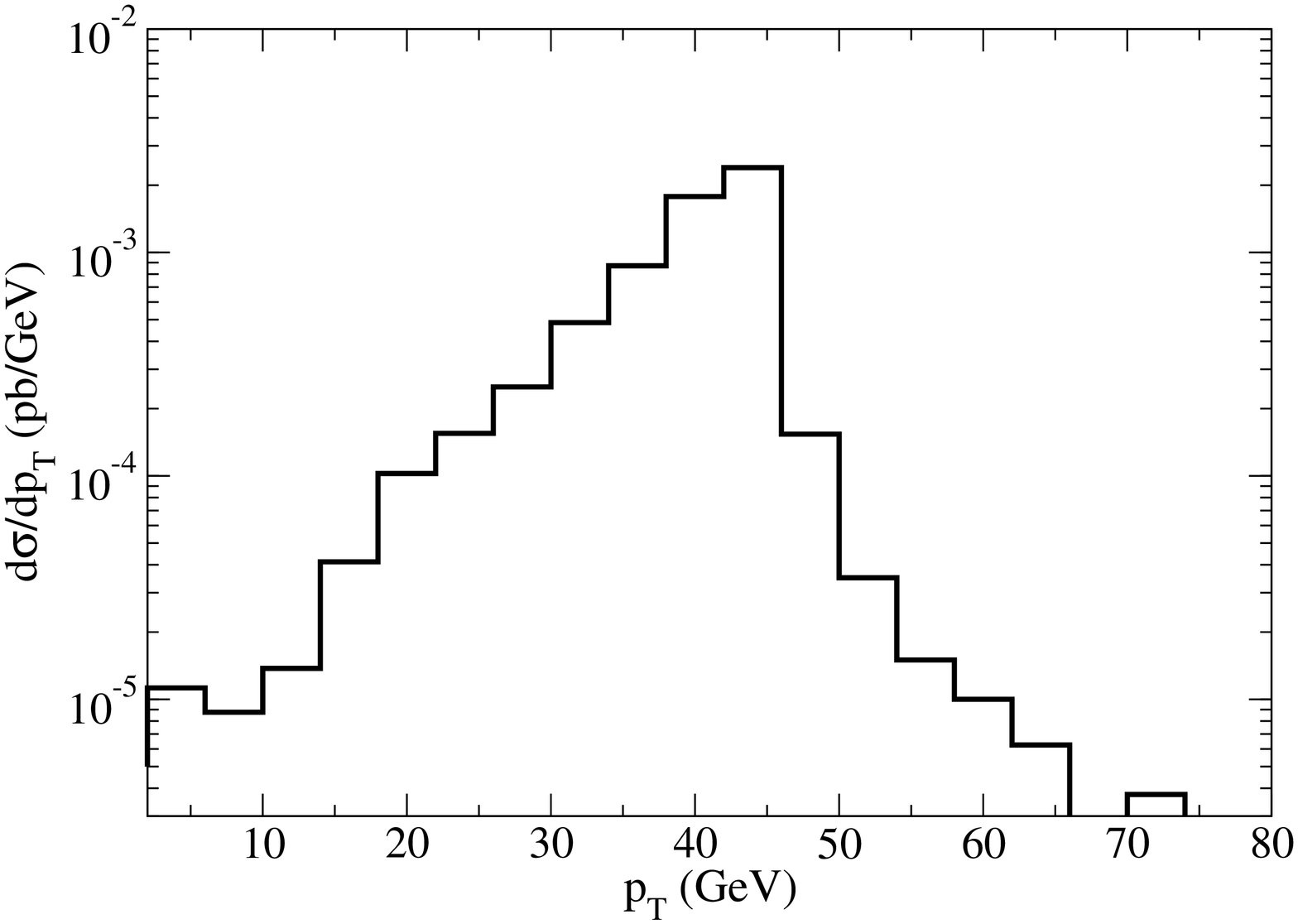}
\end{tabular}
\caption{\label{fig:z-prod}Left: The production rate as a function of the branching ratio of the decay: $Z^0 \rightarrow$ dark states. The solid and dashed lines are for $Z^0$
  decays into dark sector scalars and fermions, respectively. Right:
  lepton jet $p_T$ distribution resulting from $Z^0$ decays. }
\end{figure}

The production rates of dark sector states at the Tevatron and LHC
are shown in Fig.~\ref{fig:z-prod} \cite{Maltoni:2002qb}. We present rates coming
from decays into bosonic (denoted by $h'$) and fermionic (denoted
by $f'$) dark sector states. In the context of the SUSY models discussed later in this section,  these bosonic and fermionic states could
refer to dark Higgs bosons or Higgsinos, respectively.  On the
other hand, the collider phenomenology is similar if other
possible dark sector states decay into lepton
jets. A cut of $|\eta|<2.4$ has been imposed on the direction of the lepton jets. The difference
in rates between the fermionic and bosonic channels results from the $\eta$ cut, the boost of the $Z^0$ in the lab frame, and the fact that fermions are more likely to be emitted along the boost direction because of angular momentum conservation.

\begin{figure}[h!]
\begin{tabular}{cc}
\includegraphics[angle=270,scale=0.3]{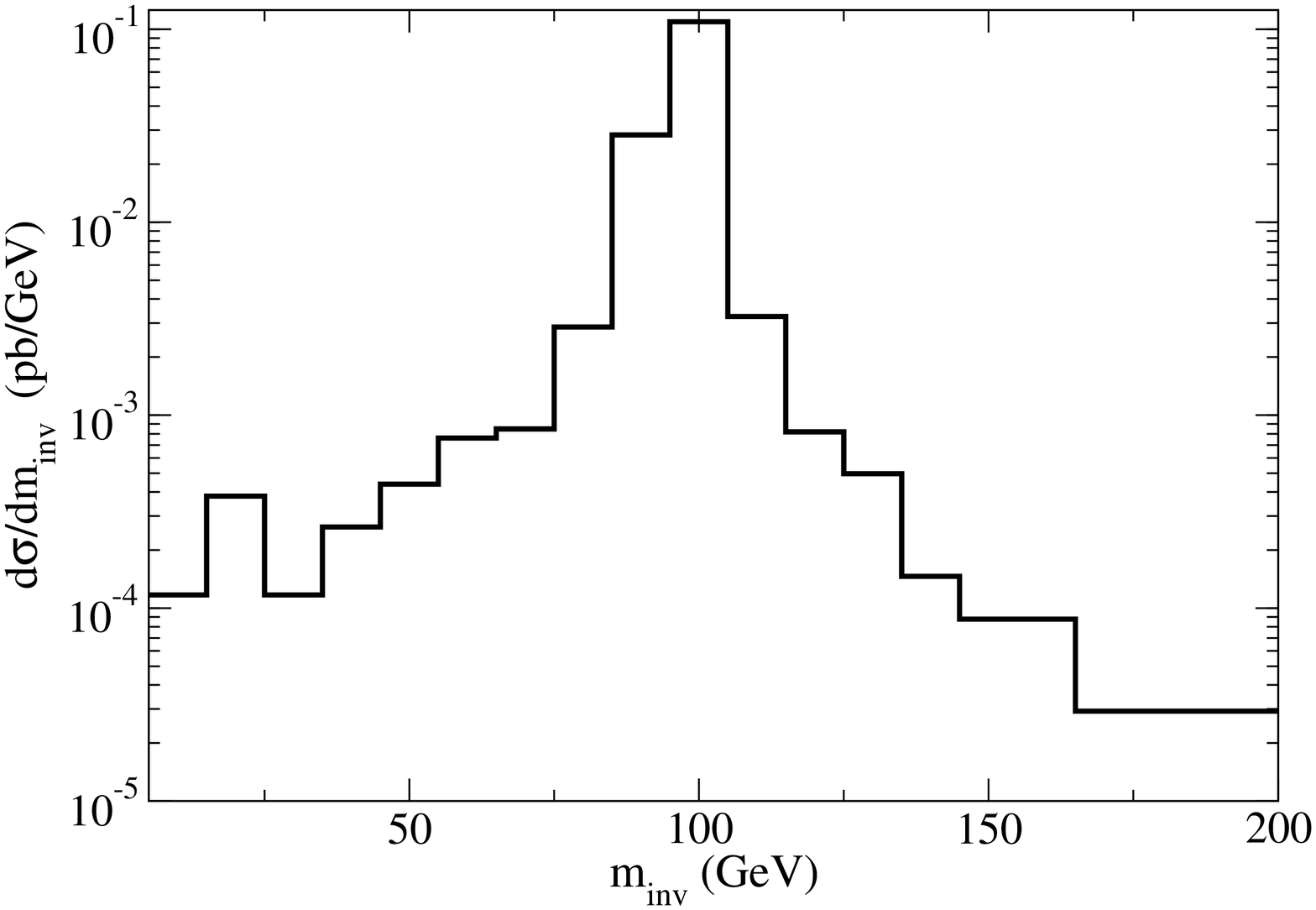}&
\includegraphics[angle=270,scale=0.3]{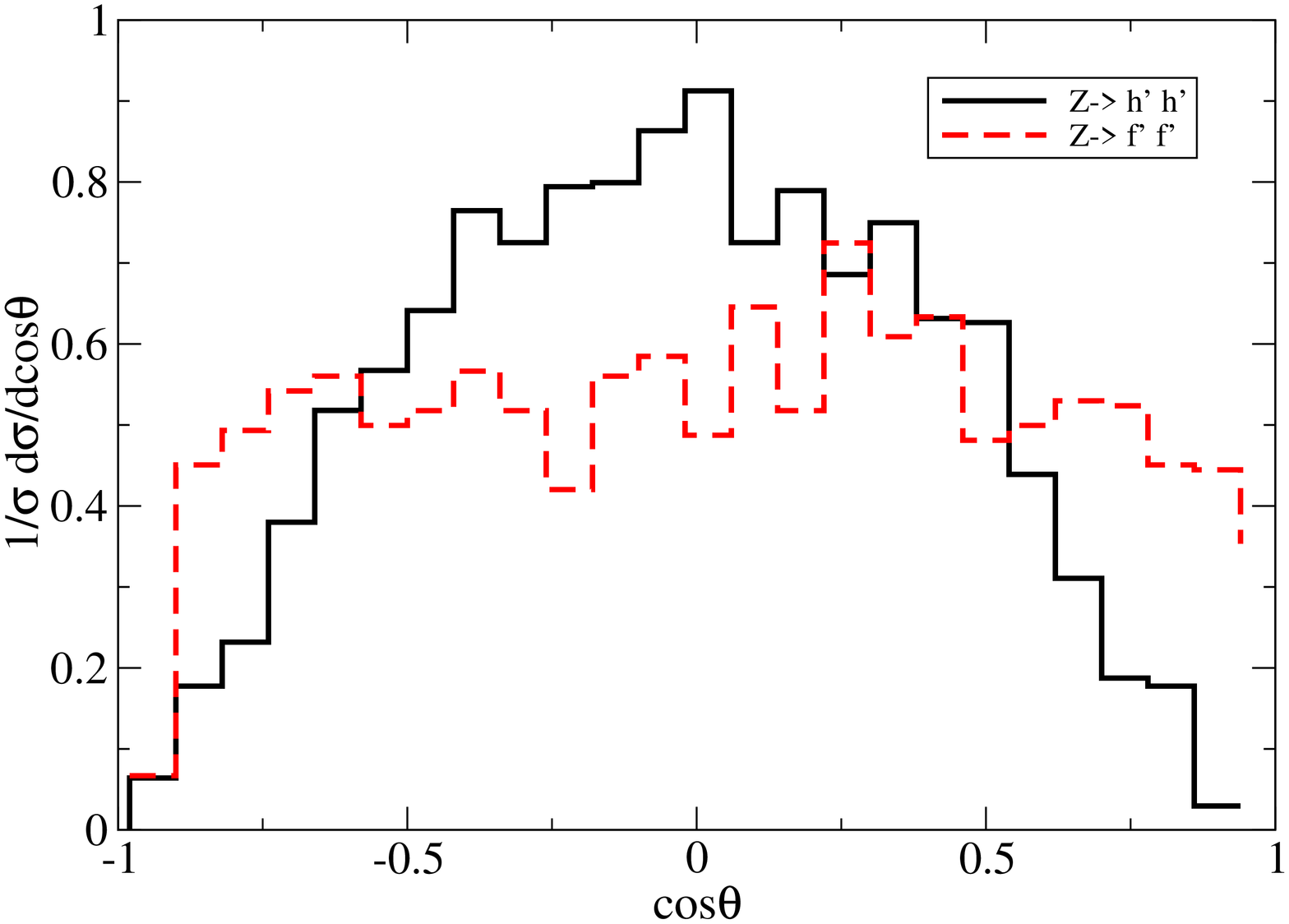}
\end{tabular}
\caption{\label{fig:z-recon-angle}Left: Reconstructed $Z^0$ boson. Right:
  Normalized lepton jet angular distribution in the $Z^0$ boson's rest-frame. }
\end{figure}

As can be seen from Fig.~\ref{fig:z-prod}, the lepton jets produced in
this way are peaked towards $p_{\rm T}^{\rm lepton \ jet} \sim 0.5 \,
M_Z$. Therefore, they are typically harder than the lepton jets resulting from
the prompt production of dark gauge bosons. As we have discussed in Section~\ref{subsubsec:direct-prod}, such harder lepton jets will be easier to trigger on. However, we expect the efficiency of identifying different leptons in a lepton jet will be lowered as it is $\propto 1/p_T$.

Reconstructing the $Z^0$ is not difficult and helps to reduce the
background. With enough statistics, it is even possible to  study
the angular distribution of the resulting lepton jets and get a
handle on the spin of the dark sector states as demonstrated in Fig.~\ref{fig:z-recon-angle}. About 5000 events are used in this plot.  We see that the expected rise for $\cos \theta \sim \pm 1$ from the fermionic decay channels is washed out due to the large boost of $Z^0$ and the $|\eta|$ cut.  However, the resulting distribution is still quite different from that of the bosonic decay channel.

\subsection{Collider signals of supersymmetric models}
In this section, we discuss the collider signals associated with supersymmetric
models. In Section \ref{sec:SUSY}, we have focused on models with
low supersymmetry breaking scale. However, the present discussion of the resulting collider signatures is largely independent of that scale or other MSSM details since we will not consider any specific superpartner spectrum.
In that sense, models with higher supersymmetry breaking scales, such as the
Planck slop option suggested in Ref.~\cite{ArkaniHamed:2008qp}, are only
different in that the gravitino is heavier. Hence, the end
of the dark sector decay chain will not involve the gravitino. However, this does not have a visible effect
on the collider signals. Even in the low
scale models where the gravitino is light, the decay length of the dark sector LSP
to the gravitino is much larger than the detector size. That said, it is important to note that the collider signatures discussed in
this section are based on the assumption that the MSSM LSP dominantly decays into the dark sector.

With supersymmetry, the dark sector states are dominantly produced
from cascade decays of MSSM colored superpartners, such as
gluinos and squarks. These particles follow typical MSSM decay
chains down to the MSSM LSP (not the gravitino). The effect of the
GeV dark sector is to extend and/or modify the decay chains
following MSSM LSP production \cite{ArkaniHamed:2008qp}. We begin
by summarizing the main features of such cascades.

\begin{figure}
\begin{center}
 \includegraphics[angle=270,scale=0.3]{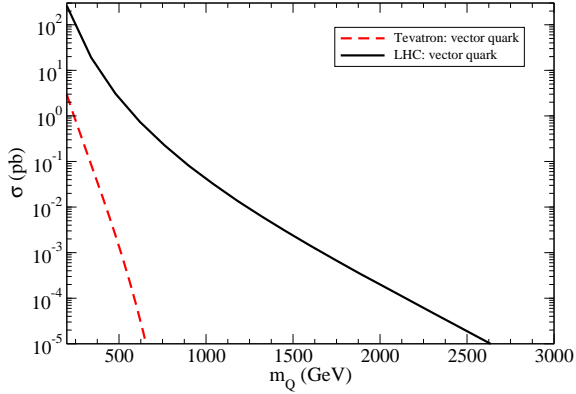}
\end{center}
\caption{\label{fig:qq_rate}Production rate of one set of $\mathbf{3}+ \mathbf{\bar 3}$ vector-like quarks, which can be part of the dark matter multiplet. The rate should be scaled by the number of such representations and the dimension of the dark matter representation under $G_{\rm dark}$. }
\end{figure}

Let us first note, however, that a notable exception occurs if dark matter is part of
a pair of $\mathbf{5} + \mathbf{\bar{5}}$ under the SM gauge groups. An example of such a model was presented in the benchmark of Section \ref{subsubsec:SUSYbench2}. The rate for the production of the colored components of such a pair is shown in
Fig.~\ref{fig:qq_rate}. Thus, the LHC has great potential for
producing such states up to about 2 TeV\@. As pointed out in
Ref.~\cite{ArkaniHamed:2008qp}, as long as the colored particles
decay only through higher dimensional operators they will be
long-lived and may have decays with very distinct signatures
\cite{Arvanitaki:2005nq}. We will not elaborate on these
possibilities but refer the interested reader to the detailed
discussion in Ref.~\cite{ArkaniHamed:2008qp}.

\subsubsection{MSSM decays into the dark sector}

Kinetic mixing implies that if the MSSM LSP is a neutralino then it decays into dark sector
states with a lifetime of
\begin {eqnarray}
\tau_{{\rm LSP} \rightarrow h+\tilde{h} } &\sim& \left( \alpha_y^{\rm dark} f_{\tilde{B}}^2 \epsilon^2  M_{\rm  LSP}\right)^{-1} \nonumber \\
&=& 7\times10^{-19}\unit{s} \;  \left(\frac{100\unit{GeV}}{M_{\rm LSP}}\right)^2 \left( \frac{0.01}{\alpha_y^{\rm dark}} \right)  \left(\frac{1.0}{f_{\tilde{B}}} \right)^2 \left( \frac{10^{-3}}{\epsilon} \right)^2,
\end {eqnarray}
where $f_{\tilde{B}}$ is the bino fraction of the MSSM LSP\@.  In
the low-scale gauge mediation models constructed earlier in this
paper, it is possible for the gravitino to be significantly
lighter than the MSSM LSP\@.  When the gravitino is light, another
possible decay channel for the MSSM LSP is LSP$\rightarrow X_{\rm SM}
\tilde{G}$, where $X_{\rm SM}$ can be a photon, Z, or Higgs, depending on
the model parameters and phase space. The decay lifetime can be
estimated as \be \tau_{{\rm LSP} \rightarrow \gamma,Z,h+\tilde{G}}
\sim \left(\frac{M_{\rm LSP}^5}{16 \pi F^2}\right)^{-1} = 3.3
\times 10^{-13} \unit{s} \; \left(\frac{100 \unit{GeV}}{M_{\rm
LSP}}\right)^5 \left( \frac{\sqrt{F}}{100\unit{TeV}} \right)^4.
\ee We see that the LSP dominantly decays into the dark sector instead
of the gravitino. However, the two channels can be comparable in
certain regions of parameter space, such as $f_{\tilde{B}} \sim
0.1$ and a low supersymmetry breaking scale $\sqrt{F}\sim \sqrt{m_{3/2} M_{P}} \sim 10$ TeV\@.

\begin{figure}
\begin{center}
 \includegraphics[scale=0.5]{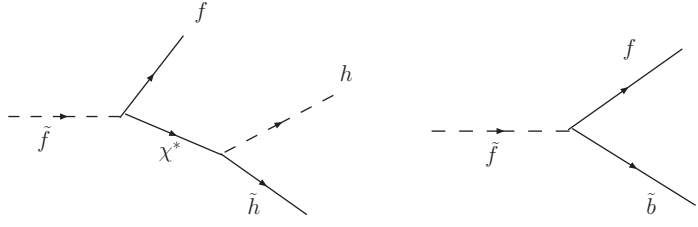}
\end{center}
\caption{\label{fig:sf_decay} Decay of sfermion LSP\@.}
\end{figure}

When the MSSM LSP is a sfermion ($\tilde{\ell}$ or $\tilde{q}$),
things become more subtle. One possible decay channel is through
an off-shell gaugino with a significant bino fraction, $\tilde{f}
\rightarrow f + \tilde{\chi}^* \rightarrow f + [\mbox{dark \
sector \ states}]$, shown the left panel of Fig.~\ref{fig:sf_decay}. Its decay lifetime can be estimated to be
\begin{eqnarray}
\tau_{\tilde{f}\rightarrow {\rm 3-body} } &\sim& \left[ \alpha_y^{\rm dark} g_Y^2 c_{f\chi}^2 f_{\tilde{B}}^2 \epsilon^2 \frac{m_{\tilde{f}}}{16 \pi^2} P( m_{\tilde{f}}/M_{\chi} ) \right]^{-1}  \\
&=& 8.3\times 10^{-16} \unit{s} \;
\left( \frac{100 \unit{GeV}}{m_{\tilde{f}}} \right)
\left( \frac{0.01}{\alpha_y^{\rm dark}} \right)
\left(\frac{1.0}{c_{f \chi} f_{\tilde{B}}} \right)^2 \left( \frac{10^{-3}}{\epsilon} \right)^2 \frac{1}{P(m_{\tilde{f}}/M_{\chi} )}, \nonumber
\end{eqnarray}
where $c_{f\chi}$ is the effective fermion-$\chi$ coupling, and $P(m_{\tilde{f}}/M_{\chi})$ is a function with the limit $P \rightarrow (m_{\tilde{f}}/M_{\chi})^4$ for $M_{\chi} \gg m_{\tilde{f}}$. Another possible decay channel is $\tilde{f} \rightarrow f + \tilde{b}$, shown in the right panel of Fig.~\ref{fig:sf_decay}, where $\tilde{b}$ is the dark bino. However, as explained in Appendix~\ref{app:KinMixing}, in addition to kinetic mixing, this coupling has an additional suppression of order $M_{\tilde{b}}/M_{\tilde{B}}$. Hence, its lifetime is
\begin{eqnarray}
\tau_{\tilde{f} \rightarrow f + \tilde{b}} &\sim& \left[ \alpha_Y \epsilon^2  m_{\tilde{f}}\left( \frac{M_{\tilde{b}}}{M_{\tilde{B}}} \right)^2 \right]^{-1}  \nonumber \\
&=& 6.6 \times 10^{-15} \unit{s} \;
\left( \frac{100 \unit{GeV}}{m_{\tilde{f}}} \right)
\left( \frac{10^{-3}}{\epsilon} \right)^2
\left( \frac{1 \unit{GeV}}{M_{\tilde{b}}} \right)^2
\left( \frac{M_{\tilde{B}}}{100 \unit{GeV}} \right)^2.
\end{eqnarray}
Notice that when the off-shell gaugino state is dominantly bino, we have
\begin{equation}
 \frac{\tau_{\tilde{f}\rightarrow {\rm 3-body} }}{
\tau_{\tilde{f} \rightarrow f + \tilde{b}} }\sim \frac{4
\pi}{\alpha_y^{\rm dark}} \frac{M_{\tilde{b}}^2
M_{\tilde{B}}^2}{m_{\tilde{f}}^4}.
\end{equation}
Therefore, these two channels can be either quite different or comparable, depending
very sensitively on the details of the model. In principle,  these
two decay channels are distinguishable experimentally, as the
three (two) body decay gives rise to three (two) different lepton
jets, respectively. Notice also that in this case, it is easier
for the channel that decays into the gravitino to be competitive
as well, if $F$ is close to tens of TeV\@.

\begin{figure}
\begin{center}
\includegraphics[scale=0.6]{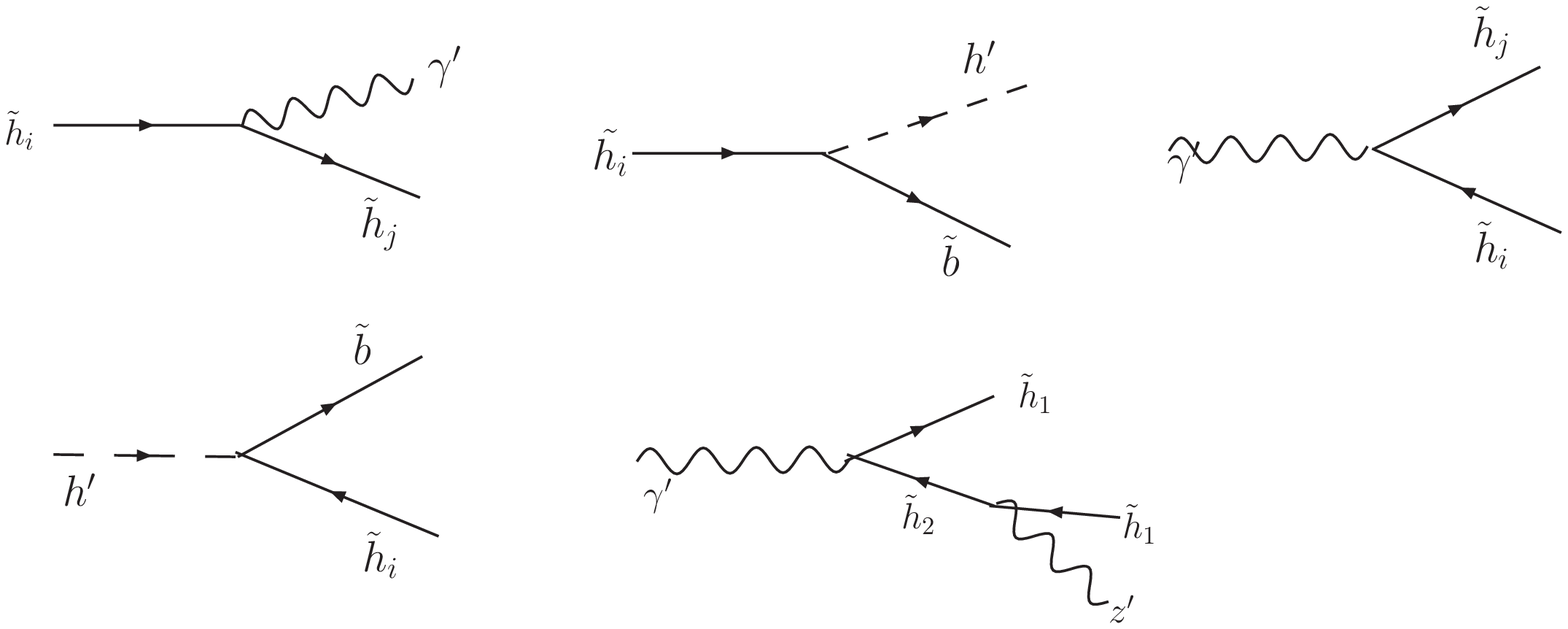}
\end{center}
\caption{\label{fig:susy-cascade} Typical SUSY dark sector decay chains. The dark sector states, $\gamma'$, $h'$ and $z'$, can be either on or off shell.  They will cascade further to produce lepton jets, similarly to the non-SUSY case.}
\end{figure}

Once the cascade has progressed into the dark sector, it will
decay through the mass hierarchy there. Several typical cascades
in the supersymmetric dark sector are shown in
Fig.~\ref{fig:susy-cascade}. For example, the center panel on the
second row is important when the fermionic dark superpartners are
lighter than the dark gauge boson. An example of such a scenario
is discussed in Section~\ref{subsubsec:SUSYbench1}. All of the
decay products in the same chain are
 collimated into one lepton jet with typical $p_T \sim 100$
GeV\@. Therefore, it is difficult to uncover the details of the
decay chain that produces a given lepton jet.

Finally, we discuss the endpoint of the dark sector decay chain.  First, consider the situation where the lightest dark sector particle is the LDSP, as in the benchmarks of sections \ref{subsubsec:SUSYbench1} and \ref{subsubsec:SUSYbench1}.
If the gravitino is heavier than the LDSP, then the decay chain will end there with the LDSP escaping
the detector, producing missing energy along the direction of the
lepton jet. If the gravitino is lighter than the LDSP, as in the
case of low scale supersymmetry breaking models, the last step of
the cascade will be LDSP$\rightarrow X_{\rm SM} \tilde{G}$, where $X_{\rm SM}$
corresponds to a light SM particle, such as a photon or lepton.
For simplicity, we consider the case where the LDSB is mostly dark
bino. We can estimate its decay lifetime as \be \label{eqn:gravitinorate}
\tau_{\tilde{b}\rightarrow\gamma\tilde{G}} \sim \left[
\frac{\epsilon^2}{16 \pi} \frac{M_{\tilde{b}}^5}{F^2}\right]^{-1}
= 3.3\times 10^3 \unit{s} \; \left( \frac{10^{-3}}{\epsilon}
\right)^2 \left(\frac{1 \unit{GeV}}{M_{\tilde{b}}}\right)^5 \left(
\frac{\sqrt{F}}{100\unit{TeV}} \right)^4, \ee which is clearly not
relevant on the collider timescale.  Second, we consider the case where there is also a dark sector gauge boson, $b$, that is lighter than the LDSP\@.  This situation is not realized in our benchmarks of Section \ref{subsec:SUSYbench}, but is certainly a possibility. 
In this case, the LDSP can decay through the channel, $\tilde{b} \rightarrow
b \, \tilde{G}$, where $b$ subsequently decays to leptons.  There is no
$\epsilon^2$ suppression here, but setting $\epsilon$ to 1 in
Eq. (\ref{eqn:gravitinorate}) gives a decay length inside the detector
only for $\sqrt{F} \lesssim 10 \TeV$.  Therefore, we can effectively
think of the LDSP as the endpoint of the dark sector decay chain for most of parameter space.

\begin{figure}[h!]
\begin{center}
\includegraphics[scale=0.5]{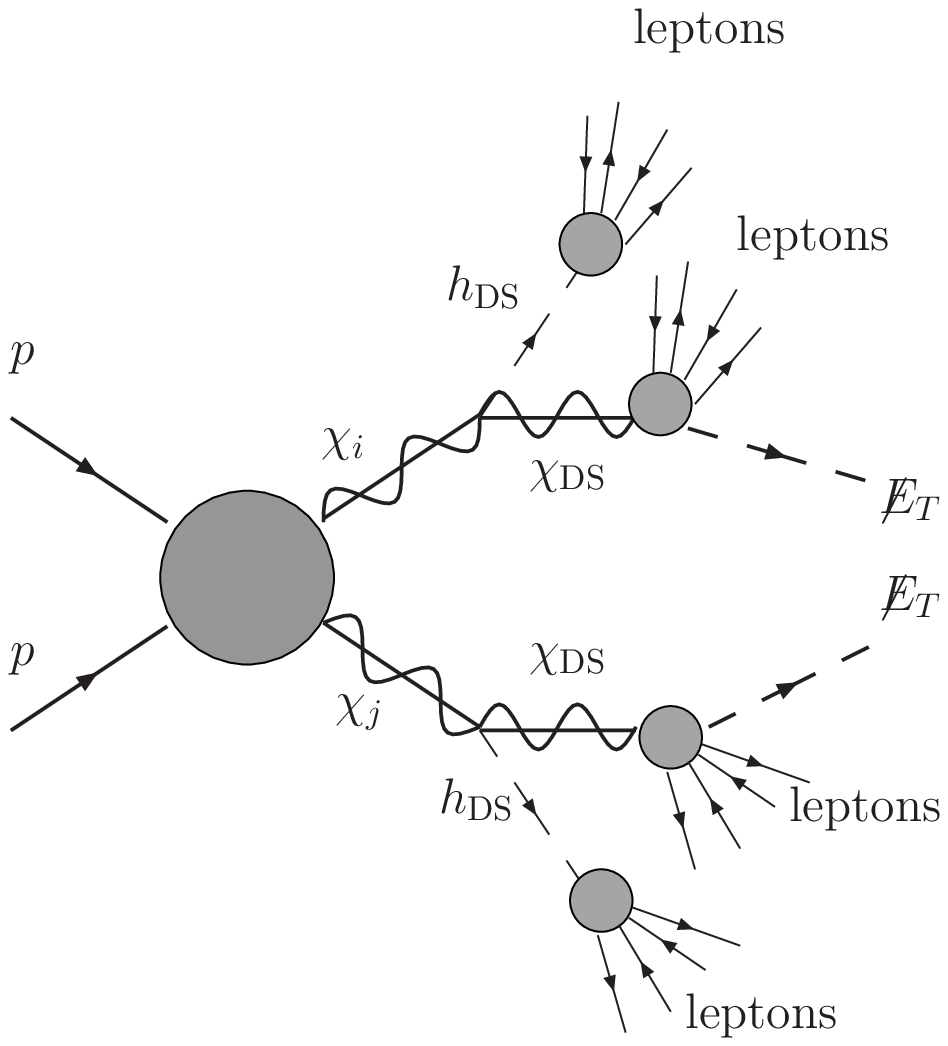}\hspace{10mm}
\includegraphics[scale=0.45]{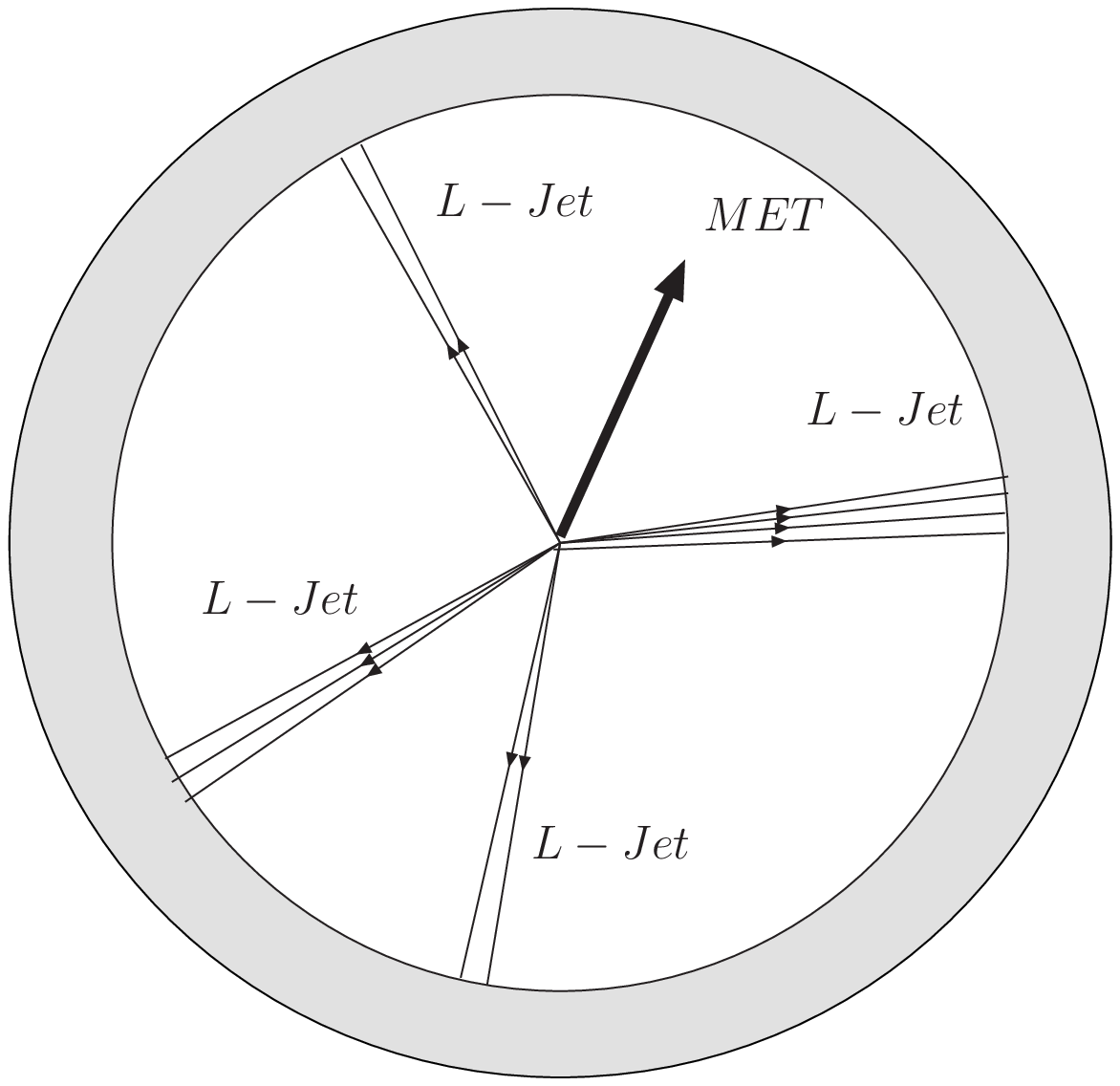}
\end{center}
\caption{\label{fig:chichiProd} Pair-production of the SM LSP can
result in spectacular lepton jet + MET events. On the left we
depict the event topology. On the right we show a schematic
representation of the resulting geometry.}
\end{figure}

\subsubsection{Extended discovery reach for direct electroweak-ino production}

\begin{figure}
\begin{center}
\includegraphics[angle=270,scale=0.25]{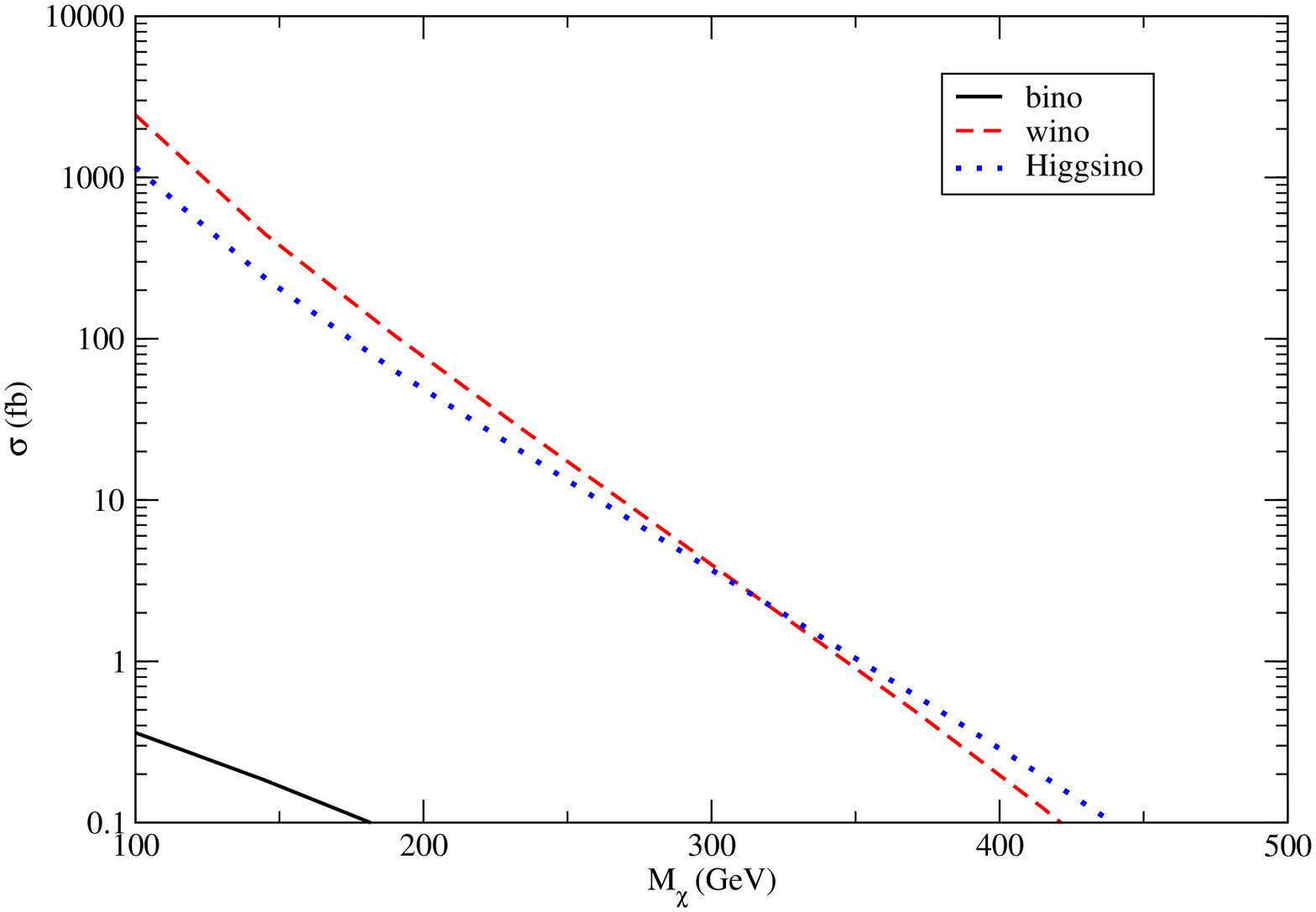}
\includegraphics[angle=270,scale=0.25]{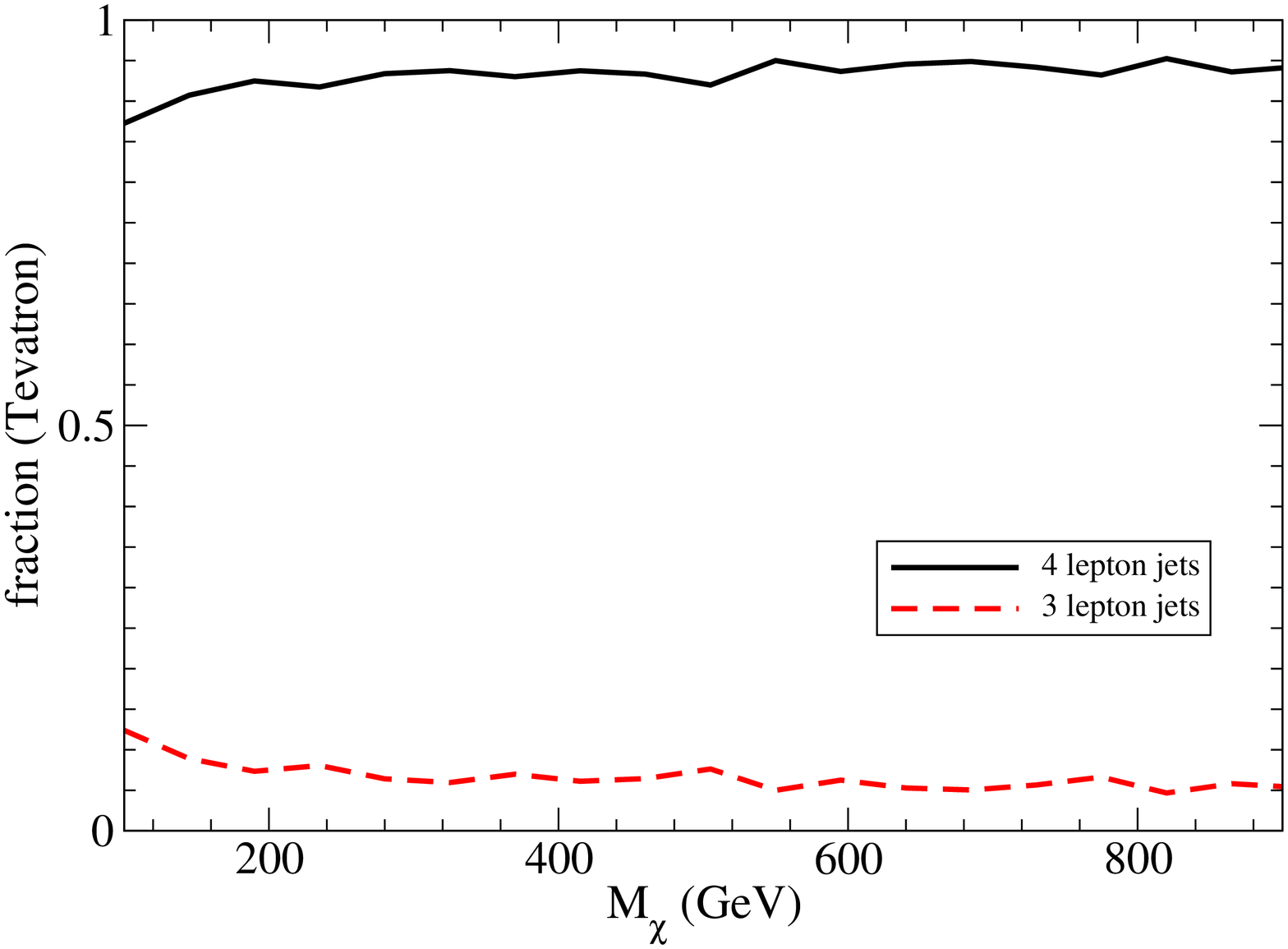}
\end{center}
\caption{\label{fig:chichiXS_TEV} Left: The cross-sections for electroweak-ino production at the Tevatron. We have included both LSP pair production and, in the case of wino and Higgsino LSP, the production of closely degenerate states, as a function of $M_{\chi}$. We choose the squark mass to be 750 GeV\@. Right: the fraction of events with 3 and 4 lepton jets within the central region $|\eta|<2.4$. }
\end{figure}

\begin{figure}
\begin{center}
\includegraphics[angle=270,scale=0.25]{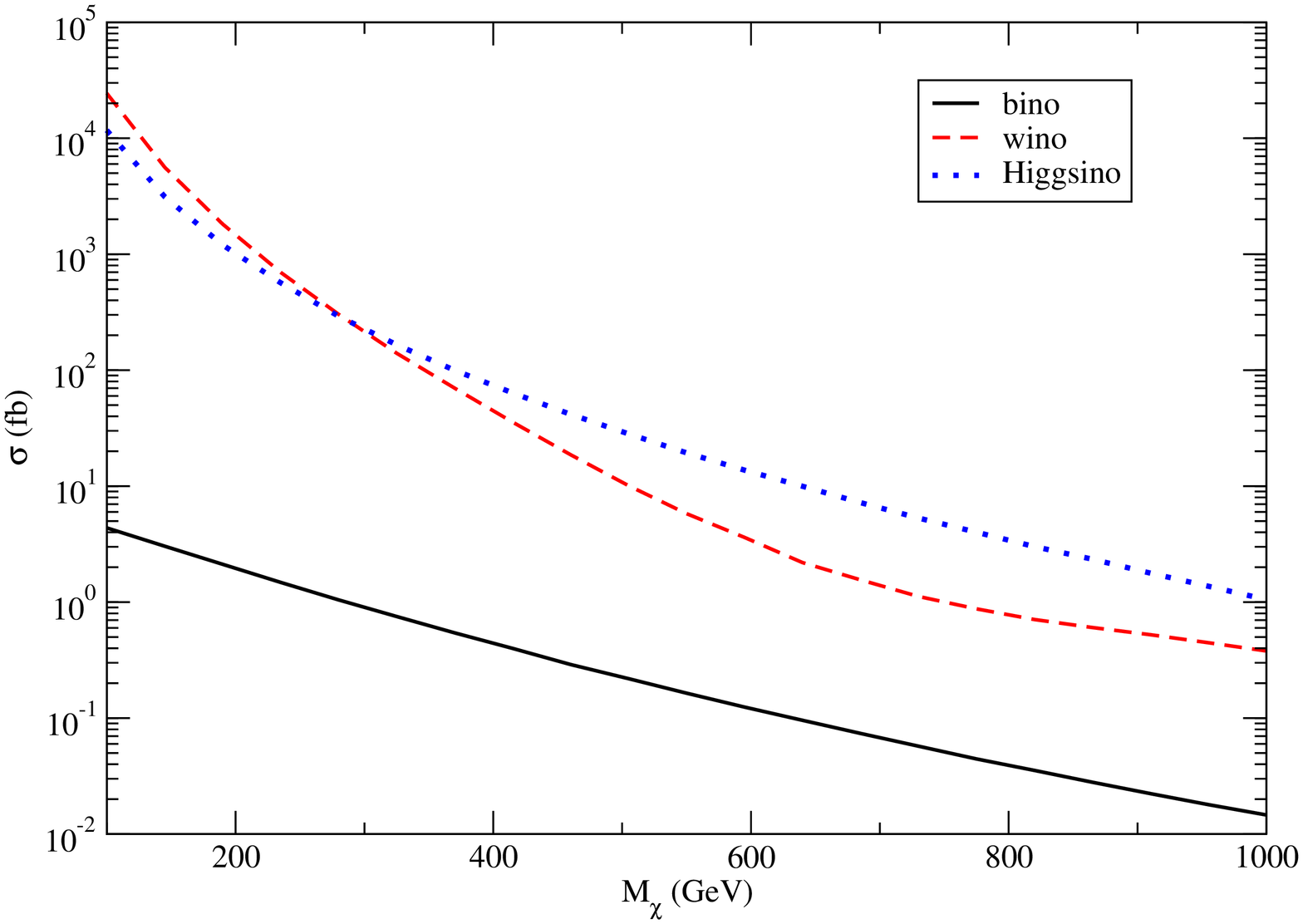}
\includegraphics[angle=270,scale=0.25]{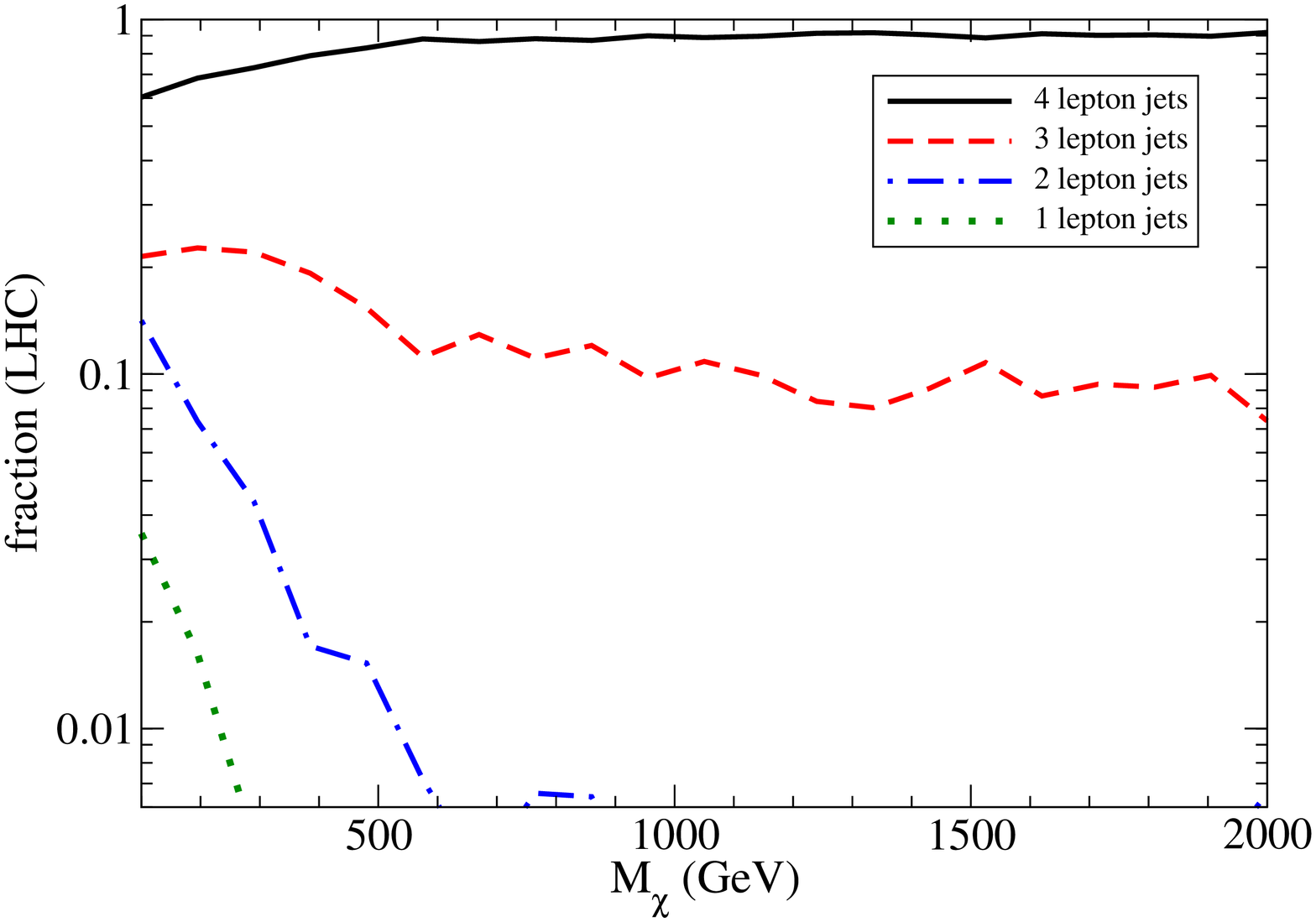}
\end{center}
\caption{\label{fig:chichiXS_LHC} Left: The cross-sections for electroweak-ino production at the LHC\@. We have included both LSP pair production and, in the case of wino and Higgsino LSP, the production of closely degenerate states, as a function of $M_{\chi}$. We choose the squark mass to be 750 GeV\@. Right: the fraction of events with 1, 2, 3 and 4 lepton jets within the central region $|\eta|<2.4$. }
\end{figure}

The direct production of electroweak-inos is an important channel since it is independent of the existence of colored superpartners and may provide additional information on the properties of those electroweak-inos. In the conventional MSSM, it is usually difficult to see events with direct pair-production of electroweak-inos.  In the case of direct MSSM LSP production, one has to trigger on
some additional hard radiation, which has a lower rate and a large
background. The pair-production of heavier electroweak-ino states which cascade down to the LSP may be easier to observe but suffers from large SM background. However, in the
scenario we consider, the LSP of the MSSM, which we denote by $\chi_{\scriptscriptstyle  0}$, will decay further into the dark states \cite{Strassler:2006qa}
whose decays result in leptons and missing energy \cite{ArkaniHamed:2008qp}. Such events
are easy to trigger on since all the leptons carry significant amounts
of $p_T$. Since $\chi_{\scriptscriptstyle  0}$ is produced almost on threshold, its boost
factor is order unity and the opening angle in the decay
$\chi_{\scriptscriptstyle  0}\rightarrow h_{\scriptscriptstyle DM}\chi_{\scriptscriptstyle DM}$ is fairly large. The resulting event geometry is striking and is depicted schematically in Fig.~\ref{fig:chichiProd}).

In the left panel of Fig.~\ref{fig:chichiXS_TEV}, we show the rate of  electroweak-ino pair production at the Tevatron.  In the case of pure wino-like and Higgsino-like LSP, we have also included the production of the closely degenerate charginos and neutralinos.  We then decay each LSP into a pair of lepton jets and study their kinematics. At the Tevatron, the neutralinos and charginos produced from $q \bar{q}'$ initial states are expected to have small boosts. Therefore, the majority of the resulting lepton jets are expected to be very central as shown in the right panel of Fig.~\ref{fig:chichiXS_TEV}, where we have required $|\eta| < 2.4$ for the lepton jets. In addition we see that the majority $> 90 \%$ of the events have 4 lepton jets within the central region as illustrated in the left panel of Fig.~\ref{fig:chichiXS_TEV}. Since the presence of such lepton jets greatly enhances the possibility of triggering on such events and separating them from the background, we estimate a reach of about $300\GeV$ for pure Higgsino or wino LSP at the Tevatron. The case of pure bino is still difficult because of the suppression in rate.

We have shown a similar study for the LHC in Fig.~\ref{fig:chichiXS_LHC}. At the LHC, the $q \bar{q}'$ initial state will carry significant boost. Therefore, as can be seen in the right panel of Fig.~\ref{fig:chichiXS_LHC}, there is a significant fraction of events with 3 or less lepton jets in the central region, especially for the smaller electroweak-ino mass $M_{\chi} \le 400$ GeV\@. On the other hand, as $M_{\chi} $ increases, the effect of the boost quickly decreases and the fraction of events with 4 lepton jets increases.  Such lepton jets will give the LHC the amazing ability to probe bino production up to $M_{\tilde{B}} \sim 1$ TeV, and wino or Higgsino production up to 2 TeV\@.

\subsubsection{Measuring the  mass of the MSSM LSP}

\begin{figure}
\begin{center}
\includegraphics[scale=0.35]{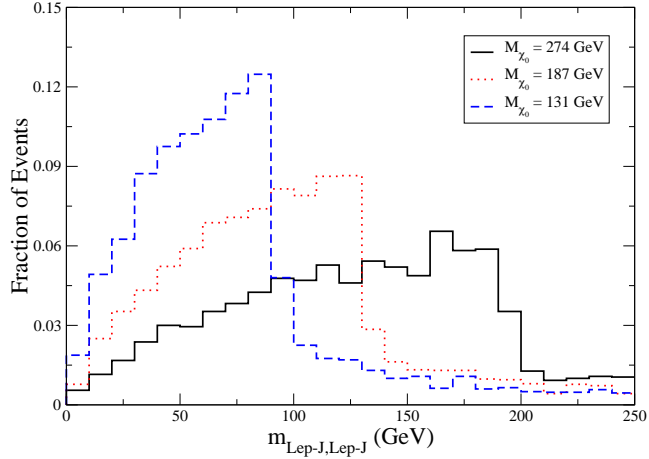}
\end{center}
\caption{\label{fig:edge} Forming the invariant mass of all the lepton jets in the events can lead to a measurement of the mass of $\chi_0$, shown here as an edge at $M_{\chi_0}$. Incorrect pairings of the lepton jets are included. However, we assumed that individual leptons are properly bunched with the correct lepton jet.}
\end{figure}

In the sorts of SUSY events shown in Fig.~\ref{fig:chichiProd}, it is possible to use lepton jets for a measurement of the mass of $\chi_0$. There are two lepton jets in each decay chain. There is a clear edge in their invariant mass distribution  at $M_{\chi_0}$, as shown in Fig.~\ref{fig:edge}. This provides an absolute mass measurement and helps to remove some of the degeneracies discussed in the literature \cite{ArkaniHamed:2005px}. In addition, such reconstructions can be very useful in other precision measurements of the properties of the MSSM superpartners. For example, since we now have information about the LSP mass, and the direction of its decay products, it is easier to reconstruct the kinematics of the full event. In fact, we can fully recover the kinematics of the event using the same reconstruction method mentioned in the case of $Z^0+$jets associated production in section~\ref{subsubsec:rareZ}.  Gaining such information will significantly improve the prospect of measuring the spin of the LSP, which can be very challenging in the conventional scenario.

\section{Conclusions}
\label{sec:conclusions}

In this paper we have explored numerous aspects of model building
and collider phenomenology for models of dark matter in which a
non-abelian dark gauge symmetry is spontaneously broken at a GeV\@.
Assuming a minimal dark gauge group, $G_{\rm dark} = SU(2) \times
U(1)$, we have surveyed a broad class of dark Higgs sectors that
break enough symmetries (charge and custodial) to be
phenomenologically viable.  Furthermore, in order to accommodate
the XDM and/or iDM scenarios, we have included the DM as the
lightest state of a gauge multiplet of $G_{\rm dark}$, which we
have shown can naturally acquire mass splittings of $\sim $ 100
keV $-$ 1 MeV, from radiative corrections and higher-dimensional
operators.

We have also argued that an attractive option is to include
supersymmetry, since this automatically generates a GeV symmetry
breaking scale for the dark sector. In particular, we have
proposed a novel mechanism whereby the $U(1)$ kinetic mixing alone
generates an effective FI term for the dark hypercharge D-term,
which in turn can induce SUSY breaking or a super-Higgs mechanism
in the dark sector at a GeV\@. We emphasize that these GeV scale
contributions will be always present, even if we choose to
communicate SUSY breaking to the dark sector via more conventional
gauge or gravity mediation. Also, in this scenario the dark matter
multiplet can be uncharged under the SM gauge group, allowing us
to evade astrophysical constraints which may arise if the dark
matter can annihilate directly to $W^{\pm}$ (which in turn decay
to $\bar{p}$ and $\pi^0$).

We have also given a detailed analysis of the gauge mediation
scenario in which the dark matter is a
$\mathbf{5}+\mathbf{\bar{5}}$ that communicates SUSY breaking from
the SM to the dark sector, as originally proposed in Ref.~\cite{ArkaniHamed:2008qp}. Several
benchmarks models with and without supersymmetry have been
explicitly constructed here as examples.

Our model building effort is far from exhaustive.  Many directions
remain to be explored. For example, in the context of
supersymmetric implementations, we have assumed a natural solution
for the $\mu / B\mu$ problems in the dark sector. More detailed
model building is certainly necessary to explore this issue
further.

We have also presented a broad analysis of possible collider
signatures of a non-abelian dark sector.  These results are
completely independent of the specific details of our benchmark
models (for example how SUSY is broken, or the exact choice of
dark gauge group). Indeed, our assumptions involved only the existence of a kinetic mixing and the non-abelian gauge symmetry of the dark sector.

We have found that dark gauge bosons can be produced via kinetic
mixing with the photon in processes analogous to prompt photon
production in the SM (with an additional suppression of
$\epsilon^2$ on the rate). At the same time, this kinetic mixing
also implies that dark gauge bosons can be produced by rare Z
decays with a branching ratio of the order $\epsilon^2$. The dark
sector states dominantly decay into SM $e^{\pm}$ and $\mu^{\pm}$.
With GeV invariant masses and typically large boosts of $\sim 10$s
GeV, these leptons typically form highly collimated ``lepton
jets''. Indeed, a key feature of a non-abelian dark sector is that
it produces lepton jets with multiple leptons, easily $>$ 2 and
possibly even 8. We expect that identifying more than 2 leptons in
a given lepton jet will significantly suppress backgrounds
and enhance signal observability.  With this in mind, there is
the promising possibility that such signals may be observed at the LHC
or possibly even at the Tevatron.  Prompt photon-like production
has a larger rate than rare Z-decay, but the latter is a
complementary process that provides more handles on the dark
sector. The 2-body decays of the dark sector into SM leptons do not generally lead to
displaced vertices. However, such displaced vertices are in general possible in the case of a 3-body decay. If the
3-body decay is further suppressed by mass splitting or if the
lightest scalar is the lightest dark state, then such a decay
leads to $\Ptmiss$ which is collimated with the lepton jet.
Several such displaced vertices will produce uncorrelated impact
parameters of decay products.

In supersymmetric implementations,
the lightest superpartner (LSP) of the MSSM will decay further
into the dark sector states which inevitably leads to lepton jets.
This offers the possibility of detecting direct electroweak
gaugino production with enhanced reach both at the Tevatron and at
the LHC\@. We can also take advantage of the unique kinematics of
such SUSY cascades to perform absolute mass measurements of the MSSM's LSP\@. Missing energy is in general present in
these events even if the gravitino is lighter than the lightest
dark superpartner (LDSP), since the decay of the LDSP$\rightarrow
\tilde{G}+X$ has a lifetime much longer than the detector timescale.

We would like to emphasize that we have only outlined the leading features of the
collider phenomenology of this scenario. More detailed simulations
are needed to obtain fully accurate estimates of the SM background with multiple collimated leptons. There are of
course variations of the models considered here which could have
new types of signals. An R-parity violating scenario could have more
intriguing decay chains into the dark sector
\cite{ArkaniHamed:2008qp}. The small coupling between the
observable sector and the dark sector could come with other types
of ``portals'', such as the ones presented in Refs. \cite{ArkaniHamed:2008qn,Nomura:2008ru},
which could lead to new collider signals. We observed that the
modification of the SUSY decay chain will in general lead to better
kinematical reconstruction. Such new information could result
in precision measurements of other properties of the
MSSM superparters involved in the decay chain. New techniques for
taking advantage of such new information are certainly worthy of
development.

\textbf{Acknowledgments}: We would like to thank Nima
Arkani-Hamed, Zohar Komargodski, David Krohn, Jim Olsen, Michele
Papucci, Maxim Pospelov, Matt Strassler, Chris Tully, and Neal Weiner for useful
discussions. Also, J. T. R. and C. C. would like to thank the
Hebrew University for their hospitality during the final stages of
this work.  L.-T. W. and I. Y. are supported by the National
Science Foundation under grant PHY-0756966 and the Department of
Energy under grant DE-FG02-90ER40542.  J. T. R. is supported by a
National Science Foundation fellowship.

\appendix

\renewcommand{\theequation}{A-\arabic{equation}}
\setcounter{equation}{0}

\section{Kinetic Mixing}
\label{app:KinMixing}

In this appendix we give a detailed description of kinetic mixing
and its effect on dark sector/SM couplings.  To begin, we consider the
non-SUSY case. Following the proposal of \cite{ArkaniHamed:2008qn}, we couple the
dark sector to the SM via a gauge kinetic mixing between the dark
and SM hypercharges (see Eq. (\ref{eq:kmix})), much like what
happens in the SM between the photon and the rho meson.  This
scheme is attractive because it does not break any symmetries of
the SM and is hence less phenomenologically constrained. Moreover,
since this operator is marginal, it can be generated at a very
high scale, and will persist in the infrared.  This implies that if both $U(1)$'s are fundamental, the kinetic mixing is a UV boundary condition sensitive to physics at the highest scales.  But if either $U(1)$ is ultimately embedded in a GUT, kinetic mixing is only induced below the GUT scale by fields charged under both $U(1)$'s.  For example, by
integrating out a multiplet of heavy fields $\Phi_i$ of mass $M_i$
that is charged under both dark and SM hypercharge, we find that
\be \epsilon &=& -\frac{g_Y g_y}{16\pi^2} \sum_i Q_{i} q_{i}
\log\left(\frac{M_i^2}{\mu^2}\right) \ee
where $Q_{i}$ and $q_{i}$ are the charges of $\Phi_i$ under dark
and SM hypercharge, and $\mu$ is the renormalization scale. If for
example, SM hypercharge is generated by symmetry breaking of some
GUT group under which $\Phi_i$ is charged, then $\sum_i Q_{i}=0$.
If this multiplet has uniform $q_{i}$ charge, then the $\mu$
dependence cancels and the argument of the log becomes some ratio
of scales in the multiplet, $M/M'$.  For reasonable sizes of $g_Y$
and $g_y$, and a log contribution $\log M/M' \sim 1$, this implies
$\epsilon\sim 10^{-4} - 10^{-3}$, which is in the right range to
explain DAMA\@.

Next, let us consider how the kinetic mixing induces couplings
between the dark sector and SM\@.  At the electroweak scale, the
terms involving the kinetic mixing are
\be \mathcal{L}_{\textrm{gauge mix}} &=& -\frac{1}{4}
W_{3\mu\nu}W_3^{\mu\nu} -\frac{1}{4}B_{\mu\nu} B^{\mu\nu}
-\frac{1}{4} b_{\mu\nu} b^{\mu\nu} +\frac{\epsilon}{2}
B_{\mu\nu}b^{\mu\nu} \\
&=& -\frac{1}{4} Z_{\mu\nu}Z^{\mu\nu}
-\frac{1}{4}F_{\mu\nu}F^{\mu\nu}-\frac{1}{4}b_{\mu\nu}b^{\mu\nu}+\frac{\epsilon}{2}(\cos
\theta_W F_{\mu\nu} - \sin\theta_W Z_{\mu\nu})b^{\mu\nu}\ee
where $F_{\mu\nu}$ and $Z_{\mu\nu}$ are the fields strengths for
the SM photon and $Z$ boson, and in the second line we have gone
from gauge eigenstate to mass eigenstate.  Performing a field redefinition on
the photon and the dark hypercharge gauge boson
\be {A_\mu}' &=& A_\mu - \epsilon \cos \theta_W
b_\mu \\
{b_\mu}' &=& b_\mu +\epsilon \sin \theta_W Z_\mu \ee
removes the kinetic mixing between the photon and $Z$, and removes the kinetic mixing between the $b$ and $Z$ up to order $\epsilon^3$.  In addition, these shifts will
modify the gauge-current couplings, $A_\mu J^\mu_{\rm em} + Z_\mu
J^\mu_Z + b_\mu J^\mu_b+ w_{\mu} J^\mu_{w}$, as well as the gauge
boson mass matrix.  Since the photon is exactly massless, the
shift of $A$ has no effect on the mass matrix and simply couples
$b$ to the electromagnetic current of the SM\@.  This is precisely
the channel that will generate the leptons seen in astrophysical
data.

Analogously, the shift of $b$ induces a coupling of the $Z$ boson
to the dark sector $b$ current.  However, unlike the photon, the
$b$ actually acquires a mass at $\sim$ GeV, and furthermore mixes
maximally with the dark $w$'s.  For this reason shifting $b$
induces a new mass mixing term between all the dark gauge bosons
and $Z_\mu$ of order $\epsilon m_b^2/m_Z^2 = (1 \GeV / 100 \GeV)^2
\times \epsilon$. This in turn generates a mass suppressed
coupling between $b$ and the $Z$ current and between $Z$ and the
$w$ current! Thus, after removing the kinetic mixing, all the
terms that couple the SM to the dark sector are
\be \mathcal{L}_{\rm coupling} &=& \epsilon b_\mu \left(\cos
\theta_W J^\mu_{\rm em}+ \mathcal{O}(m_b^2/m_Z^2)  J^\mu_Z\right)
+ \epsilon Z_\mu \left(-\sin \theta_W J^\mu_b +
\mathcal{O}(m_b^2/m_Z^2) J^\mu_w\right) \ee
where we have suppressed the mixing angles corresponding to the
higher order contributions.

If we now add SUSY, then the kinetic mixing becomes the expression
shown in Eq. (\ref{eqn:KMsuperfield}).  This induces a mixing term
for gauginos and D-terms.  Since we have already considered the
D-term mixing in Section \ref{subsect:KMBreak}, we focus on the
gauginos.  The new term is
\be \mathcal{L}_{\textrm{gaugino mix}} &=& -2i\epsilon
\lambda^\dag_{\tilde b}  \bar\sigma^\mu \partial_\mu
\lambda_{\tilde B} + {\rm h.c.}\ee
where $\lambda_{\tilde b}$ and $\lambda_{\tilde B}$ are the dark
and MSSM bino, respectively.  Once again, since the dark bino is
effectively massless at the electroweak scale, it is natural to
shift it by $\lambda_{\tilde b} \rightarrow \lambda_{\tilde b} +
\epsilon \lambda_{\tilde B}$.  This yields a coupling term
\be \mathcal{L}_{\rm coupling} &=& \epsilon \left( \lambda_{\tilde
B}
\tilde{J}_b  + \mathcal{O}(M_{\tilde b}/M_{\tilde B}) \lambda_{\tilde b} \tilde{J}_B \right) \\
\tilde{J}_b &=& g_y \sum_i q_i \tilde{h}^\dag_i h_i \\
\tilde{J}_B &=& g_Y \sum_i Q_i \tilde{H}^\dag_i H_i  \ee
where $\tilde{J}_{b}$ and $\tilde{J}_{B}$ are the fermionic
components of the dark and SM hypercharge supercurrents.  Here the
term that is $\mathcal{O}(M_{\tilde b}/M_{\tilde B})$ arises from
new mass mixing terms that arise from the gaugino shift.

\renewcommand{\theequation}{B-\arabic{equation}}
\setcounter{equation}{0}

\section{Conditions for Charge Breaking}
\label{app:CBM}
In general it is straightforward to achieve ``electroweak'' breaking
for $G_{\rm dark} = SU(2)\times U(1)$, simply by introducing a
negative mass squared at the origin of Higgs field space. In the
non-SUSY case this tachyon is inserted by hand, while in the SUSY
case it arises naturally in the gauge mediation or kinetic mixing
mediation scenarios mentioned in this paper.

However, breaking $G_{\rm dark}$ completely, i$.$e$.$ breaking
charge, is a more difficult task.  To see this, let us first
consider the two Higgs doublet model.  We can parameterize the vevs
by
\be h_1 &=& v_1 \left(
\begin{array}{c}  \cos \alpha \\ \sin\alpha \\
\end{array} \right), \qquad
h_2 = v_2 \left(
\begin{array}{c} 0 \\ 1 \\
\end{array} \right)
\ee
where $h_1$ and $h_2$ are $\bf{2}_{-1/2}$ and $\bf{2}_{1/2}$,
respectively.  For simplicity we have assumed that CP is
preserved, and we have applied an $SU(2)$ transformation to rotate
$h_2$ into a single real component. For a given Higgs potential it
is possible to determine whether charge is broken by considering
the effective potential for the charge breaking angle $\alpha$.
Charge is preserved only if $\alpha = 0$ or $\pi$ at the minimum
of the potential. Since $|h_1|^2$ and $|h_2|^2$ are independent of
$\alpha$, only two renormalizable potential terms can contribute:
$|h_1^T \epsilon h_2|^2$ and $h_1^T \epsilon h_2$.  Naively, $|
h_1^\dag h_2|^2$ contributes as well, but this term can be written
as $| h_1^\dag h_2|^2= | h_1|^2 |h_2|^2 - |h_1^T \epsilon h_2|^2$.
Expanding these contributions in terms of $\alpha$, the effective
potential becomes
\be V_{\rm eff}(\alpha) &=&  -\frac{1}{2} A \cos^2 \alpha + B \cos
\alpha \ee
where $A$ and $B$ are a function of $v_{1,2}$ and the couplings.
There is an extremum at $\alpha = \arccos B/A$. Checking that this
point is stable, we find that a necessary condition for charge
breaking is $A<0$ and $|B/A|<1$.

Next, let us consider this in an example.  In the MSSM, the
quartic couplings are fixed by the D-terms, which turns out to fix
$A=g^2>0$, which is why charge is left unbroken.  However, it is
possible to push  $A$ below zero by introducing appropriate
quartic contributions to the MSSM\@. In our SUSY benchmarks we
accomplish this by including triplets.  The one triplet SUSY
benchmark has the superpotential:
\begin{equation}
W = \mu_\Phi {\rm Tr}\left( \Phi \Phi \right) + \mu H_1^T \epsilon H_2 + \lambda H_1^T \epsilon \Phi H_2
\end{equation}
Let us consider the case where the triplet is heavy and we can
integrate it out; this yields an effective theory of doublets in
which charge breaking is simply determined.  Our SUSY benchmarks are
not in this decoupling limit, but nonetheless the physics of charge
breaking in the low-energy doublet model appears to persist even as
the triplet mass is lowered.  Integrating out the triplet yields a
quartic for the doublet with a coupling of $\lambda^2 m_\Phi^2 /
\mu_\Phi^2$, where $m_\Phi$ is the soft mass for the triplet.  Thus
in $V_{\rm eff}(\alpha)$ for this model we find
\be A &=& g^2 + \lambda^2 \frac{m_\Phi^2}{\mu_\Phi^2} \ee
We see that $A<0$ only if there is a negative soft mass for the
triplet that is appropriately large.  While this can be easily
engineered using gauge mediation, this not always possible
generically. For example, if SUSY breaking is communicated in this
theory via kinetic mixing mediation, then $m_\Phi^2$ is not
generated at the leading order, since $\Phi$ is a singlet under
the dark hypercharge.  On the other hand, we can easily remedy
this by charging $\Phi$ under dark hypercharge---however, for
anomaly cancellation we must also introduce a second triplet of
opposite charge.  In this model of two complex triplets, the
kinetic mixing mediation will generate soft masses for the
triplets.

\renewcommand{\theequation}{C-\arabic{equation}}
\setcounter{equation}{0}

\section{$\mathbb{Z}_2$ Symmetry in the $\tan \beta = 1$ Limit}
\label{app:Z2}

When $\tan \beta =1$, there is an enhanced $\mathbb{Z}_2$ symmetry
of the two Higgs doublet model that makes it phenomenologically
inviable. In particular, under this symmetry the $w_{\pm\mu}$ have
charge $-1$ and the $z_\mu$ and $a_\mu$ (dark photon) have charge
$+1$. Since only $z$ and $a$ contain a component of the dark
hypercharge, $b$, only these states couple to SM electric charge.
On the other hand, transitions between the different dark matter
states are mediated by $w_\pm$ alone. Thus it is necessary to
break this $\mathbb{Z}_2$.

The origin of this $\mathbb{Z}_2$ is as follows.  When $\tan \beta =
1$, then $h_1$ and $h_2$ have the same magnitude; thus, they can be
simply thought of as two spinors that correspond to two unit vectors
in 3-space.  Next, $h_1$ and $h_2$ uniquely define a third direction
which bisects the angle $\alpha$ between them.  Rotations of
180$^\circ$ around this axis, followed by $h_1 \leftrightarrow h_2$,
leave the vacuum invariant.  Thus, all states in the low-energy
theory are eigenstates of this $\mathbb{Z}_2$.  Since this
$\mathbb{Z}_2$ is a subgroup of $SU(2)$, it acts nicely on
$\{w_1,w_2,w_3\}$. It is obvious by choosing a basis where $w_3$
points along the axis of rotation, that two of the $SU(2)$ gauge
bosons are odd under this $\mathbb{Z}_2$ and the remaining one is
even. Thus, the latter is the only state that can mix with the dark
hypercharge, $b$, since it is also $\mathbb{Z}_2$ neutral.


\renewcommand{\theequation}{D-\arabic{equation}}
\setcounter{equation}{0}

\section{Supersymmetric Contribution to Mass Splitting}
\label{app:SUSYcont}
In this appendix we present the supersymmetric contributions to the mass corrections of a Dirac fermion, $\Psi$. We show that in the parameter region we are interested in, those are negligible. These results are well-known (see for example \cite{Thomas:1998wy}) and are presented here for completeness. We begin with the non-supersymmetric contributions. Consider, therefore, a theory with a SM-like weak gauge-group $SU(2)\times U(1)$ which in general is broken down to nothing at some scale. Also, for simplicity, we take the Dirac fermion to be charged as $\mathbf{2}_{1/2}$ with a mass $\ML$ much larger than the Higgs scale of the theory. Similar conclusions hold for any representation of the gauge-group. At low energies, the masses of the two components are split. There are both wave-functions and mass insertion diagrams, given by,
\vspace{1mm}
\begin{eqnarray}
\label{eqn:massCorrLoops}
\parbox[t]{5cm}{\vspace{-1cm} \includegraphics[scale=0.5]{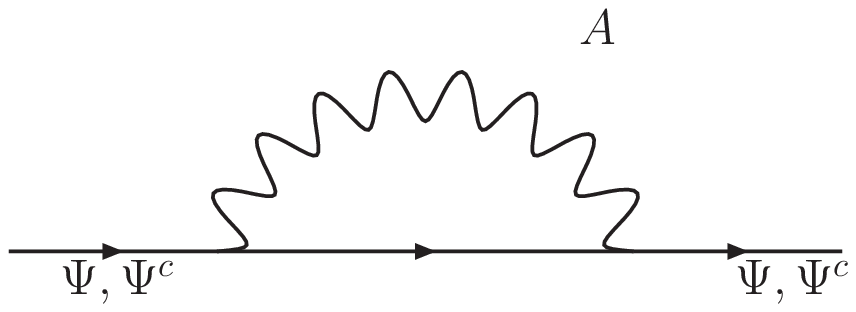} } &=& -\frac{i g_{\scriptscriptstyle A}^2}{8\pi^2}\slashed{p} \int dx x \log\Delta \\\nonumber
\\
\parbox[t]{5cm}{\vspace{-1cm} \includegraphics[scale=0.5]{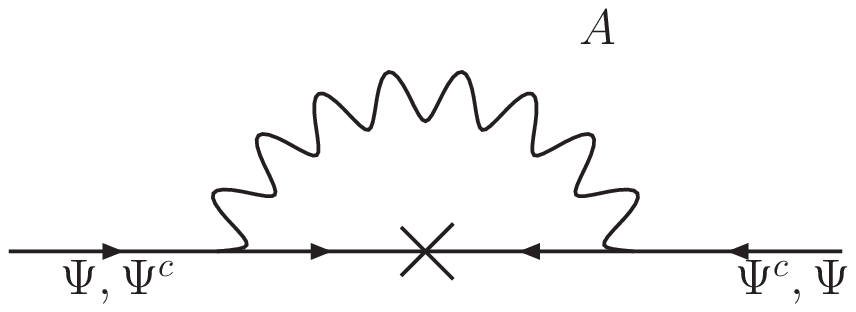} } &=& \frac{i g_{\scriptscriptstyle A}^2}{4\pi^2}\ML \int dx \log\Delta
\end{eqnarray}
where $\Delta = \left((1-x)^2 \ML^2 +x M_{\scriptscriptstyle A}^2 \right)$, $M_{\scriptscriptstyle A}$ is the mass of the gauge boson, and $g_A$ is its coupling to the fermion. We neglected the divergent part since it cancels when considering the mass splitting. The propagating gauge boson is any one of the four massive vector bosons.

For the rest of this section we consider the simple case where one gauge boson is left massless. In that case we have,
\begin{equation}
\delta \ML = \frac{\alpha \ML}{2\pi}\int dx \log\left(1+\frac{x  M_{\scriptscriptstyle Z}^2}{(1-x)^2 \ML^2}\right) \xrightarrow{\ML \gg M_{\scriptscriptstyle Z}} \frac{\alpha M_{\scriptscriptstyle Z}}{2}
\end{equation}
Where $\alpha$ is associated with the massless gauge boson
coupling to matter. For a general multiplet of the gauge group the
mass splitting between two eigenstates, $i$ and $j$ of  $T_3$ is
given by,
\begin{eqnarray}
\delta M_{ij} &=& \frac{\alpha}{2}  (q_i^2 - q_j^2)M_z \\\nonumber &-&\frac{\alpha_2}{2}\left((T^3_i)^2-(T^3_j)^2\right)\left(M_z-M_w\right),
\end{eqnarray}
where the notation is explained after Eq. (\ref{eqn:radmasssplit})
in Section \ref{sec:masssplitting}. When the ``photon'' is also
massive the correction goes as the splitting between the gauge
boson masses,
\begin{equation}
\label{eqn:nonSUSYMassSplit}
\delta \ML \approx \frac{\alpha}{2}\left(M_{\scriptscriptstyle Z} - M_{\scriptscriptstyle \gamma}\right) + \ldots
\end{equation}
However, the precise formula requires the vector boson mass eigenstates and is not simple in general.

The supersymmetric contribution is through a similar loop to the
wave-function renormalization above only with a gaugino - slepton
loop replacing the gauge boson - lepton propagators\footnote{This
way of organizing the diagrams makes it clear that even in the
supersymmetric limit, the mass insertion diagram is not cancelled
against anything else and the mass splitting is physical. This may
appear to be in conflict with known renormalization theorems of
the superpotential. However, it is important to note that such
theorems are not manifest in the Wess-Zumino gauge which is used
to compute the splitting. The splitting is a physical effect, but
the precise diagrams involved in supersymmetry is a matter of
gauge choice. For a more detailed discussion of the issues, see
Ref.~\cite{Kraus:2001kn}},
\begin{equation}
\parbox[t]{5cm}{\vspace{-1cm} \includegraphics[scale=0.5]{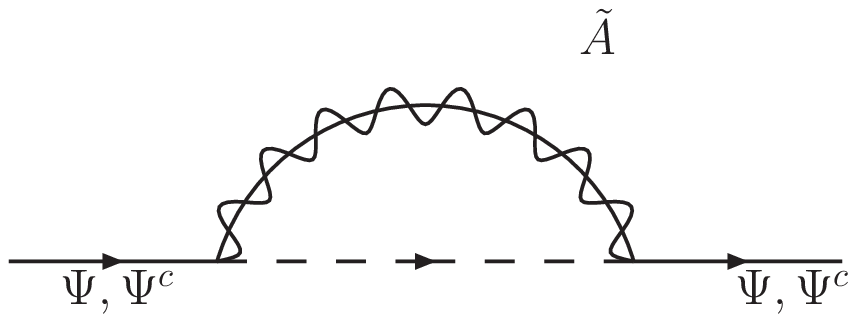} } = -\frac{i e^2}{8\pi^2}\slashed{p} \int dx (1-x) \log\tilde{\Delta}
\end{equation}
with $\tilde{\Delta} = \left((1-x)^2 M_{\tilde{\Psi}}^2 + x M_{\tilde{A}}^2 \right)$. The contribution to the splitting is then,
\begin{equation}
\label{eqn:SUSYMassSplit}
\delta \ML  = \frac{\alpha \ML}{2\pi}\int dx \log\left(\frac{(1-x)^2 M_{\tilde{\Psi}^+}^2 + x  M_{\tilde{W}}^2}{(1-x)^2 M_{\tilde{\Psi}^0}^2 + x  M_{\tilde{\gamma}}^2}\right)
\end{equation}
where we used $ M_{\tilde{W}}$, $M_{\tilde{\gamma}}$ ( $M_{\tilde{\Psi}^+}$, $M_{\tilde{\Psi}^0}$) casually to designate the charged and neutral gauge bosons (leptons). Clearly, in the limit where all the masses are equal the integral vanishes. Therefore, the only contribution to the splitting comes from possible differences between $ M_{\tilde{W}}^2$ and  $M_{\tilde{\gamma}}^2$ (or $M_{\tilde{\Psi}^+}$ and $M_{\tilde{\Psi}^0}$). If we denote the soft supersymmetry contribution to the gaugino masses by $M_\lambda$ we have,
\begin{equation}
M_{\tilde{W}}^2 - M_{\tilde{\gamma}}^2 \approx 2 M_\lambda M_W
\end{equation}
Therefore, this contribution to the mass splitting in Eq. (\ref{eqn:SUSYMassSplit}) is suppressed by $M_\lambda/\ML$ compared with the non-supersymmetric contribution in Eq. (\ref{eqn:nonSUSYMassSplit}).

Another possibility is that the charged slepton is split from the neutral one by $SU(2)$ $D$-term contributions. However, those contributions are to the mass squared. Writing $M_{\tilde{\Psi}^+}^2 - M_{\tilde{\Psi}^0} \approx M_W^2 $ we see that again the contribution to the mass splitting is suppressed by $M_W/\ML$ with respect to Eq. (\ref{eqn:nonSUSYMassSplit}).

\renewcommand{\theequation}{C-\arabic{equation}}
\setcounter{equation}{0}

\bibliographystyle{JHEP}
\bibliography{mirrorU}

\end{document}